\documentclass[journal=cmatex,manuscript=article,layout=onecolumn]{achemso}
\setkeys{acs}{keywords = true} 

\usepackage{amsmath}
\usepackage{array,longtable}
\usepackage{natbib}
\usepackage{graphicx}
\usepackage[utf8]{inputenc}
\DeclareUnicodeCharacter{2212}{\textendash}

\usepackage{longtable}
\usepackage{latexsym}
\usepackage{booktabs}
\usepackage{ctable}
\usepackage{braket}
\usepackage[caption=false]{subfig}
\usepackage{url}
\title{Oxide spinels with superior Mg conductivity}
\author{Mohsen Sotoudeh}
\email{mohsen.sotoudeh@uni-ulm.de}
 \affiliation{Institute of Theoretical Chemistry, Ulm University, 
Oberberghof 7, 89081 Ulm, Germany}
\author{Manuel Dillenz}
 \affiliation{Institute of Theoretical Chemistry, Ulm University, 
Oberberghof 7, 89081 Ulm, Germany}
\author{Johannes D\"ohn}
 \affiliation{Institute of Theoretical Chemistry, Ulm University, 
Oberberghof 7, 89081 Ulm, Germany}
\author{Julian Hansen}
\affiliation{Institute for Applied Materials, Karlsruhe Institute of Technology, Hermann-von-Helmholtz-Platz 1, 76344 Eggenstein-Leopoldshafen, Germany}
\author{Sonia Dsoke}
\affiliation{Institute for Applied Materials, Karlsruhe Institute of Technology, Hermann-von-Helmholtz-Platz 1, 76344 Eggenstein-Leopoldshafen, Germany}
\author{Axel Gro\ss}
\email{axel.gross@uni-ulm.de}
\affiliation{Institute of Theoretical Chemistry, Ulm University, Oberberghof 7, 89081 Ulm, Germany}
\altaffiliation{Helmholtz Institute Ulm (HIU) for Electrochemical Energy Storage, Helmholtzstra{\ss}e 11, 89069 Ulm, Germany}
\date {\today}

\begin{document}
\newpage
\begin{abstract}
  Mg batteries with oxide cathodes have the potential to significantly surpass existing Li-ion technologies in terms of sustainability, abundance, and energy density. However, Mg intercalation at the cathode is often severely hampered by the sluggish kinetics of Mg$^{2+}$ migration within oxides. Here we report a combined theoretical and experimental study addressing routes to identify cathode materials with an improved Mg-ion mobility. Using periodic density functional theory calculations, Mg$^{2+}$ migration in oxide spinels has been studied, revealing key features that influence the activation energy for Mg$^{2+}$ migration. Furthermore, the electronic and geometrical properties of the oxide spinels as well as their stability have been analyzed for a series of different transition metals in the spinels. We find that electronegative transition metals enable a high Mg-ion mobility in the oxide spinel frameworks and thus a favorable cathode functionality. Based on the theoretical findings, some promising candidates have been identified, prepared and structurally characterized.  Our combined theoretical and experimental findings open up an avenue toward the utilization of functional cathode materials with improved  Mg$^{2+}$ transport properties for Mg-metal batteries.
\end{abstract}

\maketitle
\newpage
 
\section{Introduction}
The field of post-lithium-ion batteries using, e.g., Na-ions~\cite{hwang17_CSR46_3529,cnaoaki14_chemrev114_11636} or Mg-ions~\cite{gregory90_jelso137_775, aurbach00_nature407_724, MacLaughlin2019, Davidson2020} has recently received substantial attention in the quest for more sustainable and abundant batteries without compromising performance parameters such as energy density and capacity. In particular the concept of  pairing a metallic anode with a high voltage cathode promises to achieve higher energy density~\cite{singh13_chemcom49_149, zhao-karger17_jmatchema5_10815} than current Li-ion technologies. Among them, batteries based on the bivalent metal magnesium theoretically reach higher volumetric energy densities compared to monovalent-based batteries.~\cite{singh13_chemcom49_149, zhao-karger17_jmatchema5_10815} However, among the materials typically used in Mg batteries only few combine an acceptable ionic conductivity with a high operating voltage. For example, MgMo$_6$S$_8$~\cite{aurbach00_nature407_724} and MgTi$_2$S$_4$~\cite{sun16_energyensci9_2273} sulfides allow an effective reversible Mg intercalation. Still, their capacities and operation potentials are too low to enable devices with a competitive energy density.

In contrast to materials based on soft anions such as sulfides, oxide materials combined with a metallic Mg anode offer high voltages and capacities.~\cite{canepa17_chemrev117_4287}. As an example, the ion-exchanged olivine-type MgFeSiO$_4$ cathode exhibits a relatively high potential of 2.4~V vs. Mg$^{2+}$/Mg and a capacity of 300~mAh/g.~\cite{orikasa14_sr4_5622} This material represents a low-cost, safe and corrosion free cathode which can be applied in rechargeable Mg batteries. Furthermore, MgMnSiO$_4$ has been considered as a potential cathode candidate material for Mg batteries.~\cite{nuli09_jpcc113_12594} However, spinel structures typcially show a lower migration barrier compared to the olivine structures. Finding spinel materials with a high ionic conductivity has therefore become an active topic of research in the field of Mg batteries.~\cite{ling13_cm25_3062, liu15_energyensci8_964, ling15_cm27_5799, kim15_advmat27_3377, dillenz20_fer8_260, Sotoudeh_Mg_Spinel, Sotoudeh_Descriptor, sotoudeh22_jpcl13_10092} 

\begin{figure}[!t]
\centering
\includegraphics[width=0.60\linewidth]{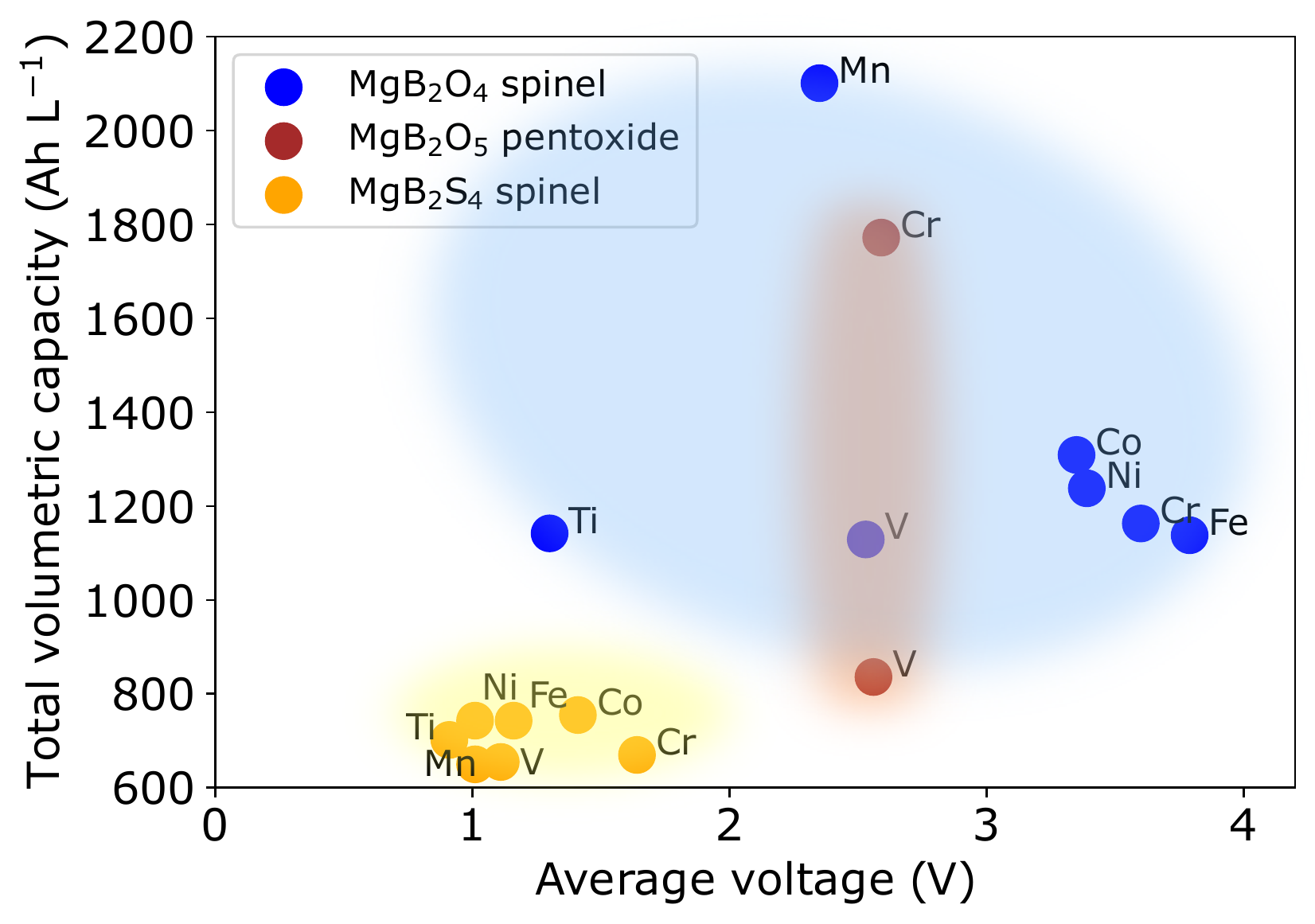}
\caption{\label{fig:fig-cap}Total volumetric capacity versus average voltage for MgB$_2$S$_4$ sulfide spinels (orange),  MgB$_2$O$_4$ oxide spinels (blue) as well as MgB$_2$O$_5$ pentoxides (brown), derived from DFT calculations within the PBE+$U$ approach. The colored areas represent the estimated range of the corresponding compounds.}
\end{figure}

As already mentioned above, inorganic oxides in the spinel structure~\cite{liu15_energyensci8_964, Bayliss20_chemmat32_663} exhibit a favorable combination of capacity and voltage, as illustrated in Fig~\ref{fig:fig-cap}.  Recently, experimental Mg$^{2+}$ migration barriers of about 600~meV have been reported in Cr- and Mn-spinel oxides, in agreement with density functional theory (DFT) calculations~\cite{Bayliss20_chemmat32_663}. It has been found that Mg-ion migration is influenced by structural disorder in the spinel lattice, which can be controlled during synthesis~\cite{Bayliss20_chemmat32_663}. However, the strong Coulombic interactions of the Mg cations with the oxide host lattice generally hamper the reversibility of Mg intercalation and lead to high Mg migration barriers~\cite{bachman16_chemrev116_140}. Recently, we have been able to identify a descriptor for the ion mobility in crystalline materials based on the ionic radii, the oxidation states and the difference in the electronegativities of the migrating cations and the anions of the host lattice~\cite{Sotoudeh_Descriptor}. Furthermore, we explored the effect of the transition metal cations in spinel compounds on their transport properties, suggesting that the transition metals can signficantly influence the Mg$^{2+}$ migration barriers in these compounds~\cite{Sotoudeh_Descriptor,sotoudeh22_jpcl13_10092}. Applying this concept, we found that oxide spinels combined with large electronegative transition metal cations promise to exhibit a rather high
Mg$^{2+}$ mobility~\cite{Sotoudeh_Descriptor}.

It should be mentioned that utilizing 4$d$ and 5$d$ transition metals instead of 3$d$ transition metals can lead theoretically to high capacity electrode materials for rechargable batteries~\cite{Yabuuchi7650, Koettgen2020}. The 4$d$ and 5$d$ transition metals had been overlooked for a long time due to their unfavorable weight. However, employing anionic redox mechanisms can enable large reversible capacities~\cite{Yahia2019}. As an example, the rock salt compound Li$_{1.3}$Nb$_{0.3}$Mn$_{0.4}$O$_2$ with the 4$d$ transition metal~Nb was found to exhibit a high reversible capacity of 300~mAh g$^{–1}$ in Li-ion batteries~\cite{Yabuuchi7650}.  Even for chalcogenide spinel lattices, it was recently shown that their Mg-ion mobility can be enhanced by selecting the heavy transition metal lanthanoid.~\cite{Koettgen2020} Therefore, the use of 4$d$ and 5$d$ transition metals in cathode materials can not only lead to high voltages but also to promising ion mobilities.

In this work, based on our descriptor concept for the ion mobility~\cite{Sotoudeh_Descriptor}, we will take the next step towards the realisation of battery materials with improved migration properties. Performing electronic first-principles density functional theory (DFT) calculations, we focus on 3$d$ oxide spinels and their Mg$^{2+}$ mobility by considering 4$d$ and 5$d$ metals as dopants in our calculations in order to improve the properties of the spinels. Based on the theoretical results, the synthesis of materials with predicted improved properties was initiated, leading to the synthesis of promising cathode materials at different doping levels. As the next step, the properties of these materials as battery cathodes will be experimentally tested. This work thus reports the first steps in the identification and utilization of promising battery materials based on an understanding of the fundamental principles underlying battery operation at the atomic level.

\section{Methods}
\label{chap:methods}
\subsection{Computational details}
First-principles calculations in the framework of density-functional theory 
(DFT)~\cite{hohenberg64_pr136_B864,kohn65_pr140_1133, Euchner2022} were performed for MgB$_2$O$_4$ spinel structures. The exchange-correlation effects were described in the generalized gradient approximation (GGA) using the Perdew-Burke-Ernzerhof (PBE) functional~\cite{perdew96_prl77_3865}. The calculations were carried out with the Projector Augmented Wave (PAW)~\cite{bloechl94_prb50_17953} method as implemented in the Vienna \textit{Ab-initio} Simulation Package~\cite{kresse93_prb47_558, kresse96_prb54_11169, kresse99_prb59_1758}. The PAW pseudopotentials include the 2$p$ and 3$s$ orbitals of Mg, the 2$s$, 2$p$ of O, the 3$d$, 4$s$ orbitals of Ti, Co, Ir, and the 3$p$, 3$d$, 4$s$ orbitals for other transition metals. 

A plane wave cutoff of 520~eV has been chosen for the wave function, and the convergence criterion for total energies was set to 1 $\times$ 10$^{-5}$ eV per supercell. A 2$\times$2$\times$2 k-point mesh has been used for the unit cell containing eight formula units with 56 atoms. The partial occupancies for each orbital are set with the tetrahedron method~\cite{jepsen71_ssc9_1763,lehmann72_pssb54_469} and the so-called Bl\"{o}chl corrections~\cite{bloechl94_prb49_16223}. The atomic positions and volume are relaxed without any restriction.

In the simulations, to properly describe the localized (strongly correlated) $d$-electrons, Hubbard $U$ corrections~\cite{Dudarev98_prb57_1505} have been used. For the 3$d$ orbitals of the transition metals, they have been set to $U_\mathrm{Ti}$ = 3.00~eV, $U_\mathrm{V}$ = 3.25~eV, $U_\mathrm{Cr}$ = 3.70~eV, $U_\mathrm{Mn}$ = 3.90~eV, $U_\mathrm{Fe}$ = 5.30~eV, $U_\mathrm{Co}$ = 3.32~eV, and $U_\mathrm{Ni}$ = 6.20~eV. For Ir and Rh, only the conventional GGA-PBE functional is used without any $U$ parameter.

The electronic structure in oxide spinels strongly depends on the type of magnetic order in the material. Both antiferromagnetic and ferromagnetic orders are probed to identify the ground state of the systems. Our calculations show that in general the ferromagnetic structures are energetically less stable than the antiferromagnetic configurations that have a zero overall magnetic moment for the 3$d$ transition metals. Thus, the electronic structure has been set to collinear antiferromagnetic spin arrangements for 3$d$ transition metals, while for Co, Rh, and Ir metals, a non-magnetic order has been assumed.

\begin{figure}[!t]
\centering
\includegraphics[width=0.60\linewidth]{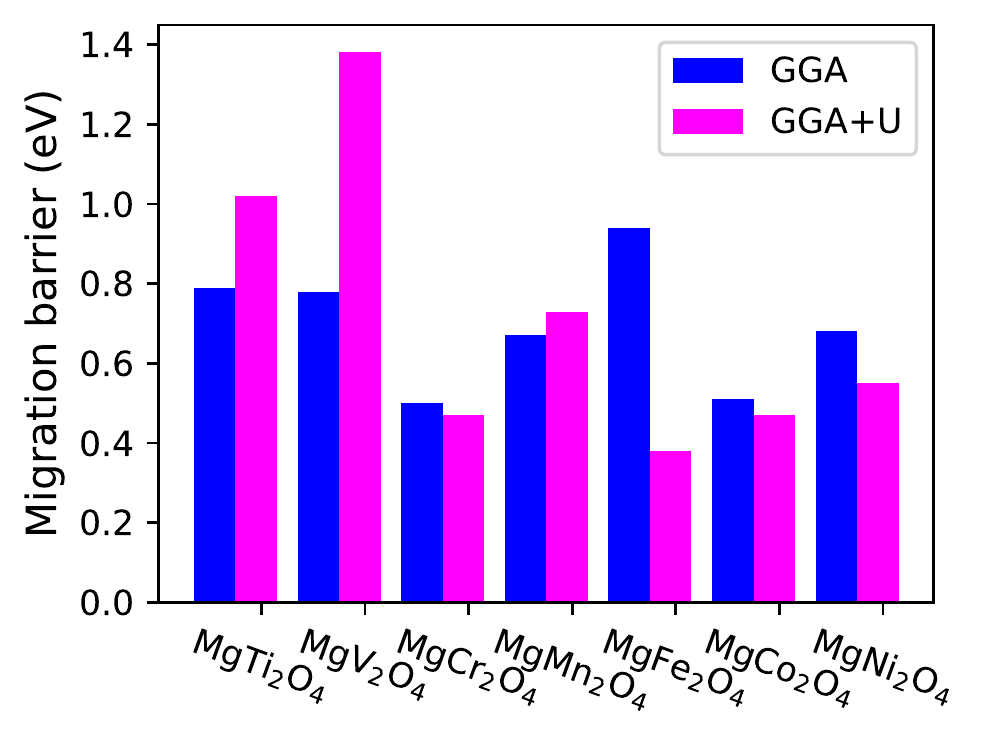}
\caption{\label{fig:gga-u}Calculated migration barrier energies using the PBE and PBE+$U$ functionals for the 3$d$ transition metal oxides in the spinel framework.}
\end{figure}

In order to identify the Mg-ion migration barriers, the nudged elastic band (NEB)~\cite{sheppard08_jcp128_134106} method was used. In order to model the Mg migration, one Mg atom was removed  within the supercell, resulting in a Mg$_{0.875}$B$_2$O$_4$ stoichiometry. It should be noted that the diffusion in the considered compounds is referred to as interstitial diffusion in the literature~\cite{mehrer2007diffusion}. During the NEB calculations, the total energies were evaluated with the PBE functional and PBE+$U$ corrections in order to validate the calculated migration barrier energies. All of the NEB calculations have been carried out with four distinct images between the tetrahedral and octahedral sites to evaluate the Mg-ion migration trajectory. The NEB calculations were fully relaxed until the forces on the atoms were converged within 0.05~eV {\AA}$^{-1}$. As Fig. ~\ref{fig:gga-u} demonstrates, including $U$ corrections severely alters the calculated migration barrier heights, for early transition metals the barriers are increased by up to 0.5 eV, whereas for the late transition metals, decreased by up to 0.5~eV. We consider the migration barriers determined with the $U$ correction to be less reliable for the following reasons:
despite the correct prediction of the ground state configurations, including all distortions, the PBE+$U$ functionals provide inaccurate energy differences between the tetrahedral site and octahedral void, namely inaccurate site preference energies, as demonstrated by comparing DFT+$U$ calculations with DFT calculations including exact exchange.  For example the site preference energy in the MgV$_2$O$_4$ is about 0.7~eV using the PBE functional and about 0.5~eV using the HSE06 functional~\cite{HSE06}, while the PBE+$U$ functional yields an energy of about -0.1~eV. Therefore, obviously the PBE functional yields a better estimate of the migration barrier energy for the considered materials in the present work than the PBE+$U$ functional. 

The  open circuit voltage ($V_{OC}$) of the spinel materials is given by
\begin{equation}
V_{OC} = -\frac{E_{inter}}{zFx}, 
\end{equation}
where $F$ is the Faraday constant, and $z$ corresponds to the elementary charges that are transferred
between anode and cathode upon the process
\begin{eqnarray}
\mathrm{Mg}_{1+y}\mathrm{B}_2\mathrm{O}_4  \rightleftharpoons \mathrm{Mg}_y\mathrm{B}_2\mathrm{O}_4 + \mathrm{Mg},
\end{eqnarray} 
with $z=2$ for Mg-ion batteries. $E_{inter}$ is the Mg intercalation energy in the spinel structure with respect to metallic magnesium which has been calculated, neglecting the zero-point energies and entropic contributions, according to
  \begin{equation}
    \label{eq:Einter}
  E_{inter} (\mathrm{Mg}) = E (\mathrm{Mg_{1+y}B_2X_4)}
         - (E (\mathrm{Mg_{y}B_2X_4}) + x E (\mathrm{Mg})) \ , 
\end{equation}
where $E (\mathrm{Mg_{y}B_2X_4})$ is the total energy of the spinel with a Mg concentration $y$ in the unit cell, and $E (\mathrm{Mg})$ is the cohesive energy of Mg bulk metal. The open circuit voltage $V_{OC}$ in volts is given by $E_{inter}/2$ for Mg-metal batteries when $E_{inter}$ is given in eV. 

Thermodynamic stability has been used to address the theoretical chemical stability and synthesizability~\cite{Bartel22_jmc57_10475}, comparing a compound's energy with all possible configurations with the same stoichiometry. The existing Materials Project~\cite{Jain13_apl1_011002} databases have been utilized to evaluate all possible decomposition products. The difference to the configuration with the lowest energy, commonly termed as the energy above hull $E_{hull}$, is provided here as a measure for stability. In order to correctly calculate $E_{hull}$, the following procedures have been taken into account: i) We have carefully checked that the formation energies of all considered compounds agree well with the corresponding values given by the Materials Project. ii) The well-known problems due to the overbinding of dioxygen in GGA calculations are accounted by a correction suggested by Wang et al. ~\cite{Wang06_prb73_195107}. iii) We do not consider the decomposition of the considered compounds into elemental metals, preventing the problem of comparing the GGA+U energies of the oxide phases with the GGA calculations of the elemental metallic phases~\cite{Jain11_prb84_045115}. In the discussion of the stability of the considered structures given below we consider structures to be stable if their energy above hull is below 0.05 eV/atom, first of all due to the fact that the accuracy of our DFT calculations is in this range, and second, because structures that are only slightly unstable with respect to the thermodynamically most stable structure might still be kinetically rather stable.

\section{Results and Discussion}
\subsection{Spinel structure}
\begin{figure}[!t]
\centering
\includegraphics[width=0.60\linewidth]{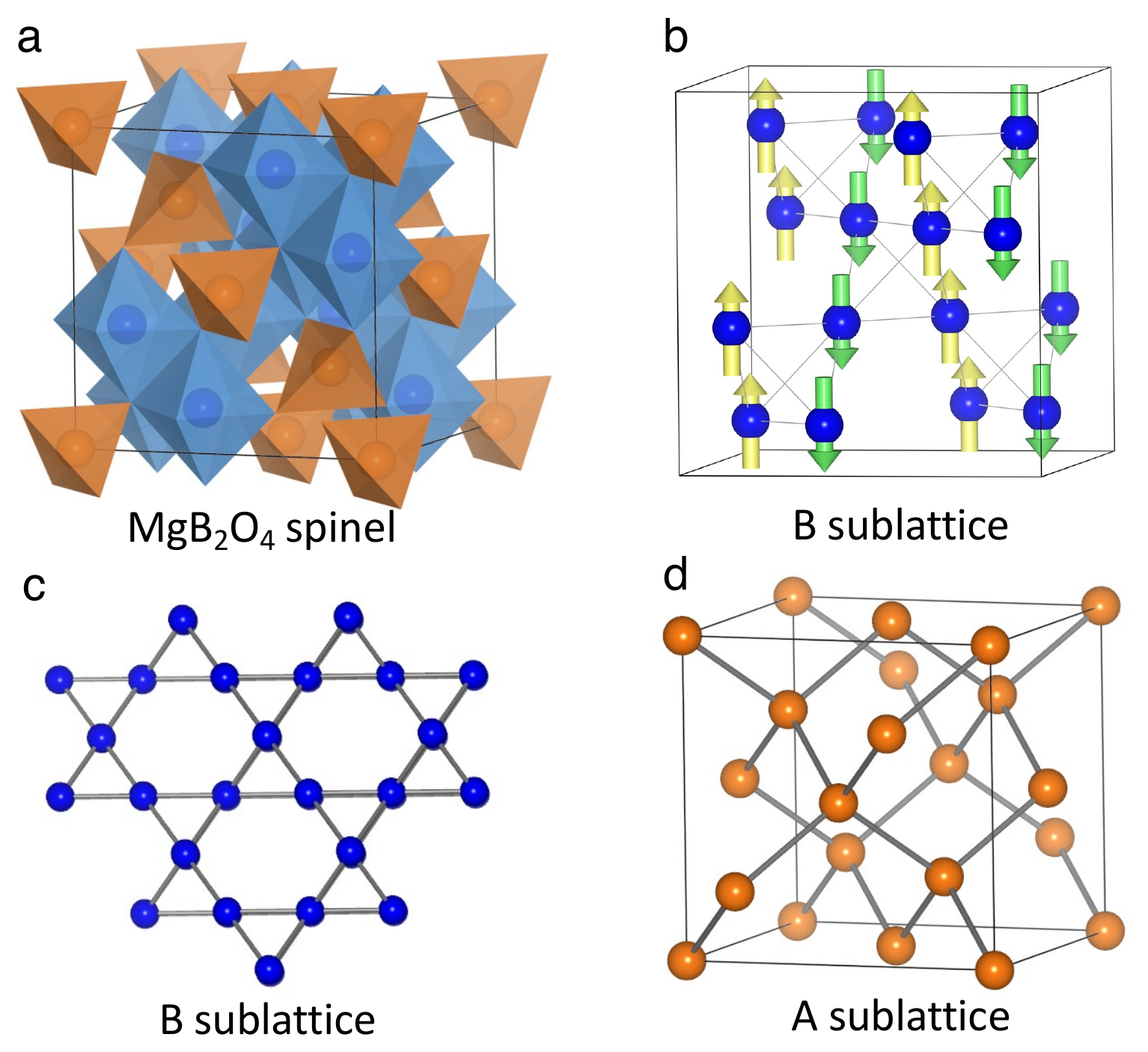}
\caption{\label{fig:fig1}(a) Spinel structure, emphasizing the two basic 
structural units of AO$_4$ tetrahedra and BO$_6$ octahedra. (b) B sublattice of 
spinel structure, which defines a pyrochlore lattice, and the lowest-energy spin 
states obtained by PBE+$U$. The B-cations are colored in blue. Green and yellow 
arrows represent the two spin directions. (c) B-sublattice structure (pyrochlore 
lattice) viewed from the $\langle 111 \rangle$ direction. (d) A sublattice 
(diamond lattice).}
\end{figure}

The oxide spinel structure which the chemical formula AB$_2$O$_4$ crystallizes with a cubic close-packed oxygen lattice. In the normal spinel crystal structure, A cations occupy $1/8$ of the tetrahedral sites, while B cations fill $1/2$ of the octahedral sites as shown in Fig.~\ref{fig:fig1}a. B and O ions form a network of edge-sharing BO$_6$ octahedra, whereas A and O  ions form a three-dimensional network of corner-sharing tetrahedra. In the oxides studied here, the size of the A-type ions is sufficiently small making their occupation of the tetrahedral site senergetically more preferable than the occupation of the octahedral sites~\cite{Sotoudeh_Mg_Spinel}. The network made by the B ions is known as the pyrochlore lattice, as shown in Fig.~\ref{fig:fig1}b. The pyrochlore lattice  gives rise to very strong geometrical frustration effects. Two types of alternate stacking planes, namely a two-dimensional triangular and a kagom\'{e} lattice, can form along the $[111]$ direction, as shown in Fig.~\ref{fig:fig1}c. The A sublattice of the spinel structure assumes the shape of a diamond lattice, which is shown in Fig.~\ref{fig:fig1}d. The presence of several competing  factors, such as the crystal field stabilization of the transition metals (B cations) or entropic contributions, can result in a variety of disordered or inverted configurations~\cite{Burdett1982}. A fraction of A cations (such as Mg) can change sites with the transition metal B cations, resulting in partially inverted lattices and charge disproportion of the B cations (2B$^{3+}$ $\rightarrow$ B$^{2+}$ + B$^{4+}$).

\begin{figure}[!ht]
\centering
\includegraphics[width=0.60\linewidth]{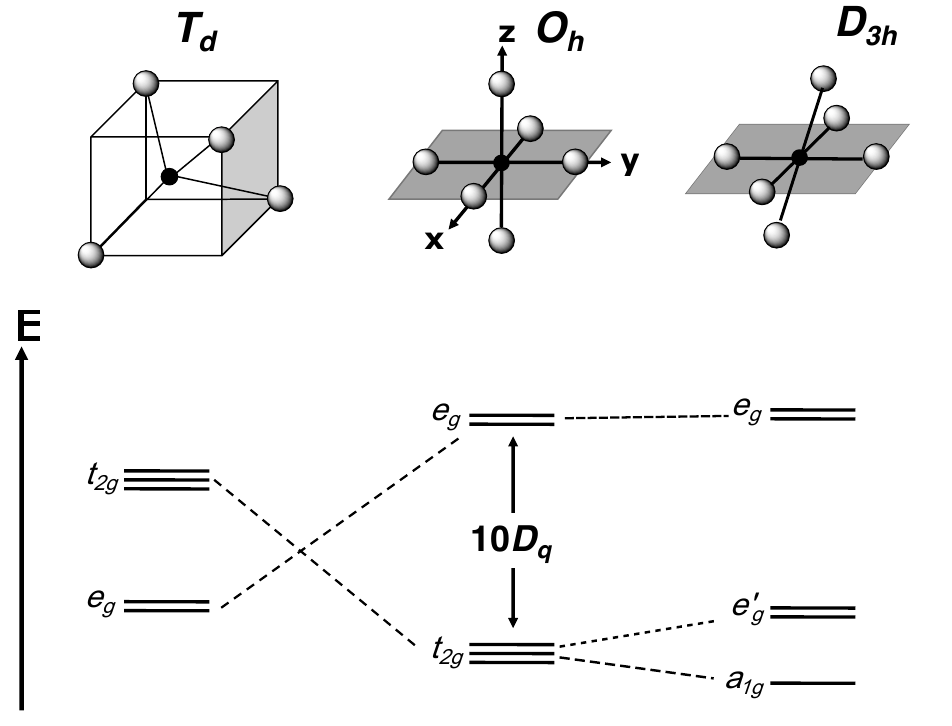}
\caption{\label{fig:sym}Schematic representation of the crystal field splitting of  $d$-orbitals at a tetrahedral site ($T_d$) and at octahedral sites without ($O_h$) and with the trigonal distortion ($D_{3h}$). The central B cation is assumed to be coordinated by oxygen atoms.}
\end{figure}

Many spinel oxides have cubic symmetry at high temperatures, but this does not rule out the possibility of distorted BO$_6$ octahedra. The octahedra undergo compression or elongation along the $[111]$ direction, pointing towards the centers of the BO$_4$ tetrahedra, as illustrated in Fig. ~\ref{fig:sym}. This gives rise to a trigonal distortion of the octahedra, changing the $O_h$ octahedral symmetry to a $D_{3h}$ octahedral symmetry~\cite{Banerjee1967}. The trigonal distortion can be parameterized by the anion parameter~$u$~\cite{Sickafus99} which reflects the displacement of the O-ions within the cubic cell. In ideal octahedra $u = 3/8$, whereas $u < 0.375$ indicates a compression of the octahedra and $u > 0.375$ corresponds to their elongation. Sickafus et al.~\cite{Sickafus99} showed that the anion parameter~$u$ can be expressed as a function of the effective radii $r$ of Mg and of the transition metal B according to
\begin{equation}
  u  =  0.3876 \left(\frac{r(\mathrm{B})}{r(\mathrm{Mg})}\right)^{-0.07054} \ .
\label{eq:u}
\end{equation}

This distortion is an important factor influencing the electronic configuration of the transition metals in the spinel structure. The resulting values for the considered oxide spinels are given in Tab.~\ref{tab:anion-parameter}. All compounds show $u$ values higher than 0.375 confirming the existence of trigonal distortion for all cases.

\begin{table}[!t]
\caption{\label{tab:anion-parameter} Calculated anion parameter $u$ 
indicating the local trigonal distortion of the BO$_6$ octahedra, using 
Eq.~\ref{eq:u} and the oxygen radius of 1.36 {\AA}. The ionic radii of the B cations 
are taken from Shannon~\cite{Shannon1976} for a sixfold coordinated configuration 
and the corresponding ionic charge of 3+ present in this list.}
\begin{center}
\begin{tabular}{lccc}
\hline
\hline
Compound  & & $r$(B)/{\AA} & Anion parameter \\
         \hline
       MgTi$_2$O$_4$     	& &  0.67		&		0.4074   \\
       MgV$_2$O$_4$      	& &  0.64		&		0.4088   \\
       MgCr$_2$O$_4$     	& &  0.62		&		0.4097   \\
       MgMn$_2$O$_4$   	& &  0.65		&		0.4083   \\
       MgFe$_2$O$_4$     	& &  0.65		&		0.4083   \\
       MgCo$_2$O$_4$    	& &  0.61			&		0.4102    \\
       MgNi$_2$O$_4$     	& &  0.60		&	    0.4106    \\
       MgIr$_2$O$_4$      	& &  0.67		&		0.4074    \\ 
       MgRh$_2$O$_4$    	& &  0.68		&		0.4070    \\ 
 \hline
 \hline
\end{tabular}
\end{center}
\end{table}

\subsection{Basic electronic configuration}
\begin{figure*}[!t]
\centering
\includegraphics[width=1.0\linewidth]{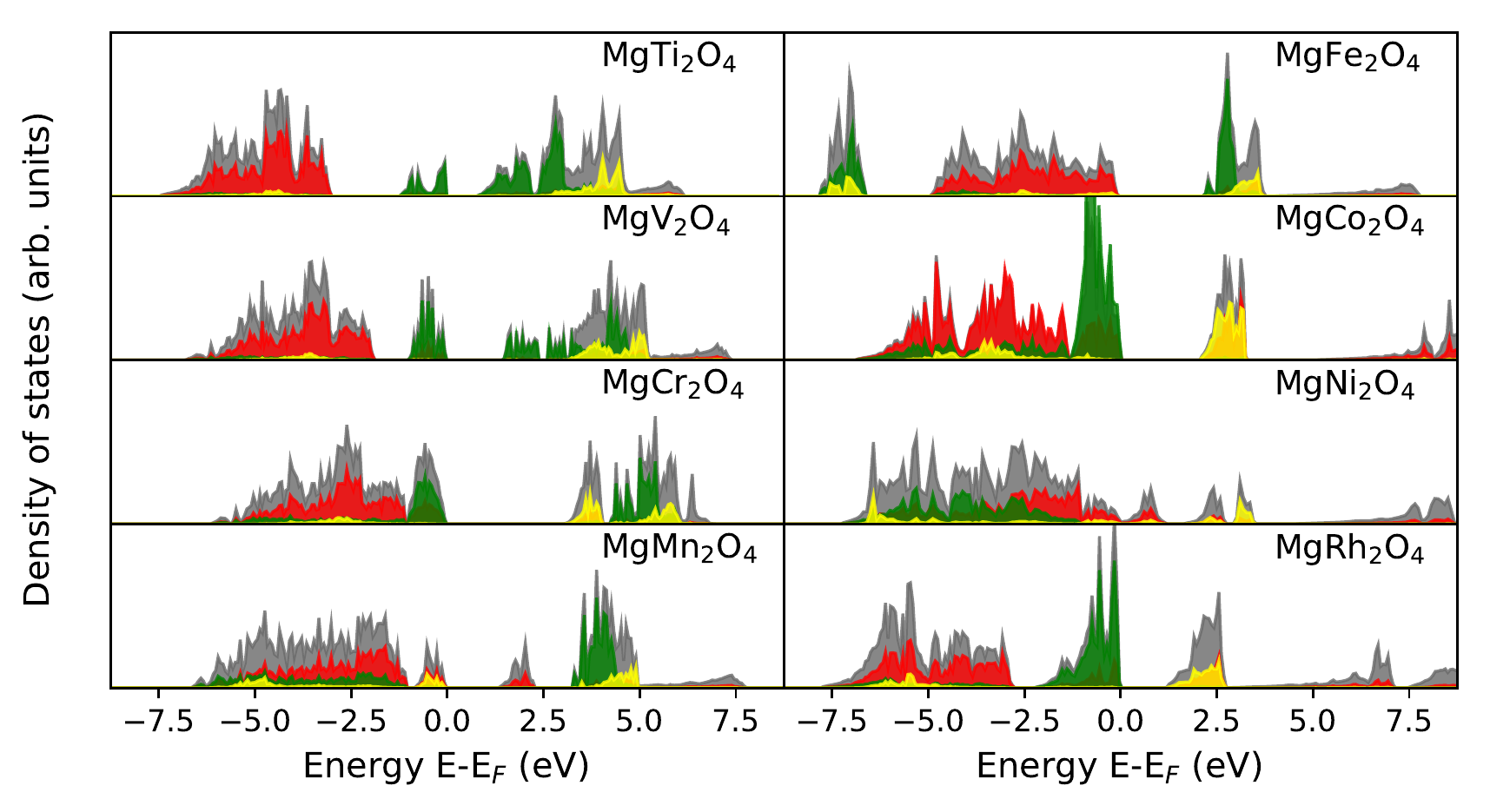}
\caption{\label{fig:dos-all} DOS of MgM$_2$O$_4$ (B = Ti, V, Cr, Mn, Fe, Co, Ni, 
and Rh) spinel structures in the stable antiferromagnetic order for 3$d$ metals 
and non-magnetic order for Rh. The graphs show the total DOS (grey) and the 
projected DOS for O-$p$ (red), B-$t_{2g}$ (green), B-$e_g$ (yellow). The 
projected DOS is considered for the B ions with both majority- and minority-spin 
directions. }
\end{figure*}

In Fig.~\ref{fig:dos-all}, we have plotted the calculated density of states whose gross features we will now summarize. Most spinel oxides containing $d$-series transition-metal ions exhibit magnetic insulator configurations due to the strong Coulomb repulsion within the $d$-orbitals. In rare cases, a spin-singlet state forms, and the system tends to be non-magnetic. Compared to sulfide spinels, oxides are less conducting due to their decreased bandwidth, which weakens  $p$-$d$ hybridization~\cite{lacroix2011magnetism}.

Crystal field splitting divides the transition-metal $d$-orbitals, located at the B site of a regular BO$_6$ octahedron, into high-lying, doubly degenerate $e_g$ and low-lying triply degenerate $t_{2g}$ orbitals, as illustrated in Fig.~\ref{fig:sym}. Note that the  crystal-field splitting originates from the covalent interaction with the oxygen neighbors. The stronger $\sigma$ bond of the $e_g$ states in comparison to the  $\pi$ bonds formed by the $t_{2g}$ states shifts the antibonding $e_g$ states energetically above the $t_{2g}$ states. In addition to the crystal field  splitting, the local trigonal distortion further divides the $t_{2g}$ orbitals  into an $a_{1g}$ orbital, and doubly degenerate $e_{g}^\prime$ orbitals. At low  temperatures, alternatively a tetragonal distortion splits the $t_{2g}$ orbitals into $xy$ and  the doubly degenerate $yz/zx$ pair and the doubly degenerate $e g$ orbitals into non-degenerate $x^2-y^2$ and $3z^2-r^2$ orbitals. 

If a transition-metal ion locates in the tetrahedral A site, the transition-metal $d$-orbitals split into the high-lying $t_{2g}$ and low-lying $e_g$ orbitals, opposite to the octahedral B-site. The B-ions in spinels usually have a large magnetic moment, leading to a Hund's rule-splitting between transition-metal $d$ levels in the majority and the minority spin direction and changes in orbital character.

\subsection{Jahn-Teller distortions}
In continuation of the preceding considerations, we now further analyze the different electronic configurations occurring in oxide spinels. The local structure around the transition metal ion will be affected by the electron filling of the 3$d$-orbitals, namely the Jahn–Teller effect. A prominent Jahn–Teller configuration is $3t_{2g}^3e_g^1$ (Mn$^{3+}$). The high-spin configuration of the Mn$^{3+}$ ion introduces a single electron in the Mn-$e_g$ state. The system lowers its energy through a tetragonal distortion which splits the doubly degenerate $e_g$ level into the singly filled $d_{3z^z-r^2}$ orbital and the higher lying empty $d_{x^2-y^2}$ orbital. In contrast, for $d^3$ and $d^5$ configurations, the $e_g$ states are either empty or filled, respectively, and therefore, no Jahn-Teller distortion occurs. The 3-fold degenerate $t_{2g}$-states form metal-oxygen $\pi$-bonds, which are weaker than the $e_g$ $\sigma$-bonds, and therefore, the Jahn-Teller effect is smaller compared to the $e_g$-states. Hence, weak distortions are expected fo $3d^1$ (Ti$^{3+}$), $3d^2$ (V$^{3+}$), and $3d^3$ (Cr$^{3+}$) as well as the $3t_{2g}^4e_g^2$ (Co$^{3+}$) and $3t_{2g}^5e_g^2$ (Ni$^{3+}$) configuration. The Rh- and Ir-spinels show no Jahn-Teller distortions due to the predominant ferromagnetic order of these spinels which enhances the $p$-$d$ hybridization.

\subsection{Selected oxide spinels}
After reviewing the gross features of the atomic and the electronic structure of the oxide spinels, we will now discuss how transition metals affect the Mg migration in the series of magnesium oxide spinels considered in this study. We concentrate on normal spinel phases that crystallize in cubic single-phase lattices (space group $Fd\overline{3}m$, see Fig.~\ref{fig:fig1}).

\subsubsection{MgTi$_2$O$_4$}
The first member of the class of oxide spinels considered in this study is MgTi$_2$O$_4$. The ions are in the formal oxidation states Mg$^{2+}$Ti$_2^{3+}$O$_4^{2-}$. Thus, an electron is inserted into the $t_{2g}$ states of the Ti sites. The 3$d^1$ electronic configuration together with the total spin $S = 1/2$ make this system to a strong candidate for the occurrence of dimerization in a single-valence system. In fact, MgTi$_2$O$_4$ undergoes a metal-insulator (M-I) transition when cooled below $T_\mathrm{M-I} = 260$\,K, accompanied by a strong decrease of the magnetic susceptibility and a transition to a tetragonal structure. In the tetragonal structure of MgTi$_2$O$_4$, Ti moves away from the center of the TiO$_6$ octahedron, and the six nearest-neighbor Ti-Ti distances become inequivalent. 

The calculated DOS of MgTi$_2$O$_4$ is shown in Fig.~\ref{fig:dos-all}. The filled valence band, which extends from -7 eV to 0 eV, consists of predominantly O-$p$ character from -7 eV to -3V with some contribution of Ti-$d$ orbitals from -1 eV to 0 eV. The unfilled Ti-$d$ states are located 1-5 eV above the valence band. The Ti-based spinel has the lowest voltage with respect to a Mg metal anode among all the Mg spinel compounds based on other redox-active cations. The Ti-based compound has an energy of 0.013\,eV/atom  above \textit{hull} which shows the stability of the compound and the possibility to be synthesized. Compared to an ideal octahedron that possesses Ti-O distances with equal length, two Ti-O bonds are compressed by 6\% in this system. Thus, the octahedron has both tetragonal and trigonal distortions, within the corresponding point group $D_{3h}$. 

The Ti oxide spinel is associated with a low open-circuit voltage with respect to magnesium metal as well as rather high Mg migration barrier, i.e. a low ion mobility, as detailed in Tab.~\ref{tab:strc}. The inverse spinel, Mg$_2$TiO$_4$, is more stable, which makes Mg-ion migration even more difficult. Thus this 
compound is not suited as a cathode material for Mg batteries.

\begin{table*}[!htb]
\caption{\label{tab:strc}Mg-O, and B-O  bond lengths in {\AA} for spinel 
compounds. B denotes the transition-metal. Calculated relative barrier energy $E_a$, 
energy above hull $E_{hull}$, intercalation energy $E_{inter}^{high}$ 
($E_{inter}^{low}$) (Eq.\,\ref{eq:Einter}) for high (low) Mg concentration in 
eV, and corresponding open-circuit  voltage $V_{OC}^{high}$ ($V_{OC}^{low}$)
with respect to a Mg metal anode in V. The volume changes with respect to 
the structure without Mg is indicated by $\Delta V/V$.}
\begin{center}
\resizebox{\textwidth}{!}{%
\begin{tabular}{lccccccccc}
\hline
\hline
Compound  & Mg-O ({\AA}) & B-O ({\AA}) & 
$E_a^{PBE}$(eV) & $E_{hull}$(eV/atom)
& $E_{inter}^{high}$ (eV) & $V_{OC}^{high}$ (V) & $E_{inter}^{low}$ (eV) & $V_{OC}^{low}$ (V) & $\Delta V/V$ (\%) \\
         \hline
       MgTi$_2$O$_4$     & 2.027 & 2.025/2.147   & 0.791 & 0.013  &-2.258 & 1.129 & -3.805 & 1.903 & -3.5\\
       MgV$_2$O$_4$      & 1.993 & 2.040/2.072  & 0.783 & 0.013  &-5.464 & 2.732 & -5.090 & 2.545 & -1.4\\
       MgCr$_2$O$_4$    & 1.994 & 2.031/2.033   & 0.497 & 0.037  &-7.318 & 3.659 & -6.994 & 3.497 & -1.5\\
       MgMn$_2$O$_4$   & 1.998 & 1.961/2.312    & 0.672 & 0.036 &-5.958 & 2.979 & -4.817 & 2.409 & -0.9\\
       MgFe$_2$O$_4$    & 2.002 & 2.049/2.053  & 0.940 & 0.000 &-7.696 & 3.848 & -6.959 & 3.479 & 0.8\\
       MgCo$_2$O$_4$    & 1.965 & 1.931			   & 0.513 & 0.039 &-6.754 & 3.377 & -8.688 & 4.344 & 3.2\\
       MgNi$_2$O$_4$     & 1.973 & 1.878/2.030   & 0.680 & 0.064  &-7.188 & 3.594 & -4.311 & 2.155 & -2.0\\
       MgIr$_2$O$_4$      & 2.064 & 2.096             & 0.319 &  0.287   &-3.956 & 1.978 & -4.557 & 2.278 & 3.6\\ 
       MgRh$_2$O$_4$    & 2.027 & 2.069             & 0.429 & 0.000  &-4.390 & 2.195 & -5.871 & 2.935 & 2.4\\ 
 \hline
 \hline
\end{tabular}
}
\end{center}
\end{table*}

\subsubsection{MgV$_2$O$_4$}
Passing from MgTi$_2$O$_4$ to MgV$_2$O$_4$, the octahedral site is now filled by V$^{3+}$ which accommodates two electrons into the triply degenerate $t_{2g}$ orbitals. The orbital degrees of freedom of these electrons allow a more complex interplay of frustrated spins and orbitals. The trigonal distortion plays a significant role in this compound while the tetragonal distortion is not pronounced. The Vi-O bonds are compressed by 1\%, splitting the $t_{2g}$ orbitals.  

MgV$_2$O$_4$ is a S = 1 Mott insulator with a pyrochlore lattice. The DOS of MgV$_2$O$_4$ shown in Fig.~\ref{fig:dos-all} reveals that filled V-$d$ states are located above the oxygen valence band in the range from -1 eV to 0 eV. The empty V-$d$ states are located in the range from 1.5 eV to 6 eV. Compared to the Ti spinel, the $t_{2g}$ orbitals are filled to a larger degree and are shifted downward close to the O-$p$ states.  Similar to the Ti spinel, this compound has an energy of 0.013 eV/atom above \textit{hull} indicating that this material could be synthesized and it may be stable.

As Tab.~\ref{tab:strc} shows, the calculated open-circuit voltage for the V spinel is about 2.7~V, however, the height of the migration barrier for Mg is 0.783~eV, indicative of a low ion mobility. The normal spinel, MgV$_2$O$_4$, is more stable than the inverse one. However, inspite of the high potential, the low ion mobility prohibits the use of this compound as cathode material.

\subsubsection{MgCr$_2$O$_4$}
MgCr$_2$O$_4$ is an S = 3/2 Mott insulator with trivalent Cr. The three electrons are accommodated in the triply degenerate $t_{2g}$ orbitals of Cr-$d$, which all three degenerate states singly occupied, and cooperative distortions of the CrO$_6$ octahedra are absent with no orbital degrees of freedom.  This compound exhibits a geometrical frustration associated with the pyrochlore structure of the Cr-sublattice. 

The calculated DOS of MgCr$_2$O$_4$ shown in Fig.~\ref{fig:dos-all} exhibits filled Cr-$t_{2g}$ orbitals located in the upper part of the O-$p$ band in the range from -1\,eV to 0\,eV. In contrast to the Ti and V spinels, the filled $t_{2g}$ orbitals are merged with the upper part of the oxygen valence band. The bottom part of the conduction band consists of Cr-$e_{g}$ orbitals in the majority spin orientation. Structurally the Cr-O bonds are almost equal in the CrO$_6$ octahedra, and solely trigonal distortions play a role. This compound has an energy above \textit{hull} of 0.037 eV/atom which indicates that the structure is possibly stable. Compared to the other 3$d$ oxide spinel, it has relatively small Mg migration barriers.

The small migration barrier of 0.497~eV and open-circuit voltage of 3.6~V demonstrate that this compound can potentially be used as a cathode material (see Tab.~\ref{tab:strc}). The reported experimental value of 0.62$\pm$0.10~eV for the Mg migration barrier in MgCr$_2$O$_4$  at high temperature~\cite{Bayliss20_chemmat32_663} is consistent with our calculation. 

\subsubsection{MgMn$_2$O$_4$}
MgMn$_2$O$_4$ is an S = 2 Mott insulator with four electrons in the Mn-$d$ orbitals. The $t_{2g}$ orbitals are filled by three electrons, and an additional electron is inserted into the anti-bonding $e_g$ states of the Mn sites. The presence of the electron in the $e_g$ orbitals causes the formation of Jahn-Teller polarons. Structurally, two Mn-O bonds in the MnO$_6$ octahedra are elongated by 18\% and cause a splitting of the doubly degenerate $e_g$ orbitals. Thus, the bandgap in MgMn$_2$O$_4$ is affected by this Jahn-Teller effect. Note that the weakening of the Mn-O antibonding orbitals with neighboring oxygen atoms upon the increase of the corresponding Mn-O bond distances causes a stabilization of the occupied states.

As shown in Fig.~\ref{fig:dos-all}, the filled valence band in MgMn$_2$O$_4$, which extends from -6 eV to 0 eV, is predominantly of O-$p$ character with some contribution from the Mn-$d$ orbitals. The filled Mn-$e_g$ states in the majority spin directions are located in the upper part of the oxygen valence band in the range from -1 eV to 0 eV. The empty Mn-$e_g$ states in the majority spin direction are in the range from 1.5 eV to 2 eV. In this compound, both tetragonal and trigonal distortions play a signficant role. The energy above \textit{hull} of 0.036 eV/atom indicates possible stability of the structure. 

Furthermore, we obtained a migration barrier of 0.672~eV in agreement with the experimental value of 0.690$\pm$0.090\,eV~\cite{Bayliss20_chemmat32_663}.Still, compared to the other 3$d$ oxide spinel, this only allows a relatively low mobility for the Mg ions due to the high migration barrier. The open-circuit voltage is 2.98~V. Nonetheless, the rather low stability makes the formation of the Mg-Mn antisite defects upon inversion and consequently an even larger migration barrier very likely. It should be mentioned that MgMn$_2$O$_4$ stabilizes in the tetragonal crystal structure at low temperature. Thus a phase transition occurs  from the cubic structure to the tetragonal one. Still, compared to the other 3$d$ oxide spinel, the Mn spinel shows relatively high voltage and stability, which makes it a potential candidate for cathode usage. 

\subsubsection{MgFe$_2$O$_4$}
MgFe$_2$O$_4$ is an S = 5/2 Mott insulator with half-filled Fe-$d$ orbitals. As Fe$^{3+}$ accommodates five electrons in the $d$ orbitals within the high spin configuration, this system exhibits a geometrical frustration of the Fe-sublattice with no orbital degree of freedom. The band gap lies between filled O-$p$ and the empty $t_{2g}$ states on the Fe site in the minority-spin direction which determines all important features of the transport properties in the Fe-spinel. Structurally, the Fe-O bonds in the FeO$_6$ octahedra are elongated by 0.5\%, which, however, causes no Jahn-Teller active mode due to the small distortion. 

In contrast to the other 3$d$ spinel oxides, the top of the valence band has an O-$p$ character,  and the Fe-$d$ orbitals in the majority spin direction split off below the O-$p$ valence band from -6.5~eV to -8~eV, as shown in Fig.~\ref{fig:dos-all}. The conduction band consists of Fe-$d$ states in the minority-spin direction in the range of 2-4~eV. The bandgap is about 2~eV predominantly in majority-spin direction. All Fe-O bond lengths are equal showing that only trigonal distortions play a role in this compound. The small energy above \textit{hull} of 0.000 eV/atom indicates a high stability which means that one should be able to prepare this material. However, this compound shows the highest Mg migration barrier and thus the lowest Mg-ion mobility compared to the other considerd 3$d$ oxide spinels. Note that the Fe spinel lattice has a half-closed shell which presumably makes it challenging to remove electrons.

The Fe spinel possesses the highest open-circuit voltages among the considered oxide spinels, as detailed in Tab.~\ref{tab:strc}. However, despite the high voltage, this compound suffers from low ion mobility.

\subsubsection{MgCo$_2$O$_4$}
The Co spinel oxide is a non-magnetic Mott insulator in which its electrons in the Co-$d$-orbitals are paired. This compound accommodates six electrons in the $d$-orbitals, introducing three electrons with opposite signs in the $t_{2g}$ states. Thus, this system exhibits no geometrical frustration in the Co-sublattice, orbital, and spin degrees of freedom. Structurally, the Co-O bonds in the CoO$_6$ octahedra are identical, and only the trigonal distortion plays an important role. 

The calculated DOS of MgCo$_2$O$_4$ (see Fig.~\ref{fig:dos-all}) shows a filled valence band which extends from -7 eV to 0 eV, predominantly of O-$p$ character with the filled contribution of Co-$t_{2g}$ orbitals in the upper part of the valence band. The conduction band dominantly consists of the empty Co-$e_{g}$ contributions   along with oxygen bands. In this compound, the trigonal distortion is strong compared to the other considered oxide spinel and causes a broadening of the Co-$t_{2g}$ orbitals.
 
This Co oxide exhibits a good ionic conductivity reflected in the relatively low Mg migration barrier of 0.513\,eV, hence it might be a good candidate as a cathode material. The energy above \textit{hull} of 0.039\,eV/atom shows that the spinel framework is may be the stable phase for Co. The Co spinel shows a relatively high open-circuit voltage of 3.38~eV which makes it a promising candidate as electrode material. However, among the considered oxide spinels, this material has a relatively large volume change of 3\% upon charging/discharging which is, however, still acceptable (see below).

\subsubsection{MgNi$_2$O$_4$}
In contrast to the other normal spinels, MgNi$_2$O$_4$ is weakly metallic with S = 3/2. This compound accommodates seven electrons in the $d$-orbitals, introducing two electrons in the minority-spin direction of the $t_{2g}$ states. There is an interplay of frustrated spins and orbital degrees of freedom in this system. Structurally, two Ni-O bonds are elongated by 8\% in the NiO$_6$ octahedra which leads to the presence of distortion modes.

The calculated DOS of MgNi$_2$O$_4$ in Fig.~\ref{fig:dos-all} shows a continuously filled valence band from -7 eV to 0 eV, predominantly of O-$p$ and Ni-$d$ character. At the valence band top, the Ni-$t_{2g}$state in the minority-spin direction splits into filled and empty orbitals with a small gap. Thus, the small bandgap in MgNi$_2$O$_4$ is due to the splitting of Ni-$t_{2g}$ in the minority-spin direction. This compound shows a strong trigonal distortion compared to the other considered oxide spinel, but also tetragonal distortions play a role.

The energy above \textit{hull} of 0.064 eV/atom indicates a low possibility for synthetization and stability of the material. However, this compound has a rather large Mg migration barrier indicating a low ionic conductivity. The calculated open-circuit voltage is about 3.6~eV, which is relatively high. Our calculations demonstrate a drop in potential to about 2.1~eV  for low Mg concentrations. Furthermore, the tetragonal crystal structure is more stable, as Mg prefers the octahedral sites in this compound.

\subsubsection{Mg(Rh/Ir)$_2$O$_4$}
We turn now to oxide spinels with 4$d$ and 5$d$ transition metals located in the B sites. These compounds often show a ferromagnetic ordering of the spins. The identical magnetic direction of the spins does not induce any distortion and the B-O bonds show no elongations or compressions. Electronically they are not Jahn-Teller active.

It should be noted that in spinel oxides, the B-O bonds are predominantly covalent. We have previously shown~\cite{Sotoudeh_Descriptor} that the migration barrier depends sensitively on the B-O bond length and the electronegativity difference between the B and O atoms. 

\begin{figure}[!t]
\centering
\includegraphics[width=0.60\linewidth]{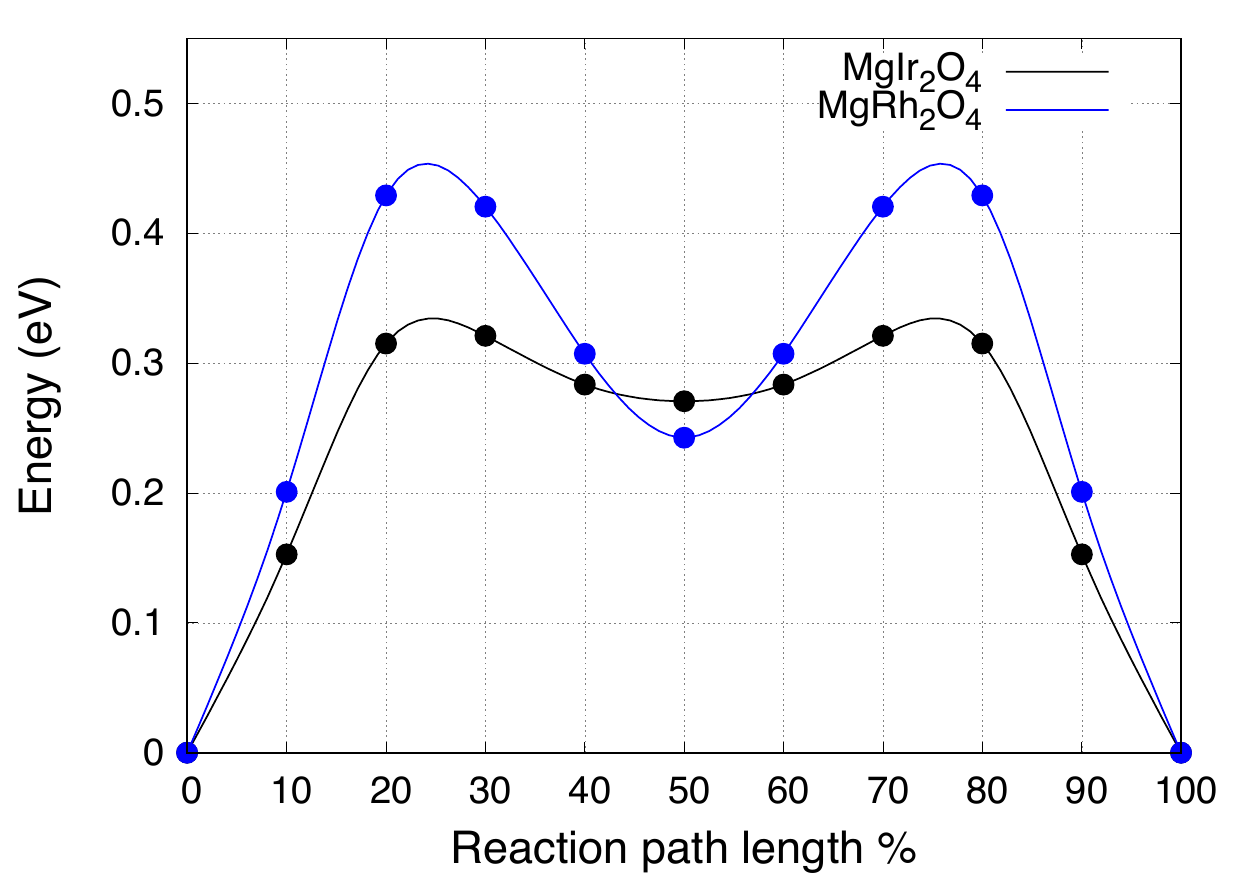}
\caption{\label{fig:neb-Ir/Rh}Mg migration energy barriers (in eV) as a function of the reaction path coordinate derived from periodic DFT calculations combined with NEB for the single-ion migration from tetrahedral site to the octahedral void and then to the next tetrahedral site corresponding to the MgIr$_2$O$_4$ spinels with the black color and the MgRh$_2$O$_4$ spinels with the blue color. The minimum energy is set to zero.}
\end{figure}

We found a minimum in the height of the migration barriers for electronegativity difference values of $\Delta{\chi}^2$ $\approx$ 2. Based on the identification of this pronounced minimum and the corresponding matching properties of Rh and Ir, we identify MgIr$_2$O$_4$ as a super-ionic conductor with high Mg mobility with a migration barrier of only 0.319~eV, as shown in Fig.~\ref{fig:neb-Ir/Rh}. Rh with a similar electronegativity as Ir also leads to a high Mg mobility in the spinel oxide framework with a migration barrier of only 0.429~eV (see Fig.~\ref{fig:neb-Ir/Rh}). 

The electronic structure of these two compounds is similar to each other. They are non-magnetic oxides, solely with trigonal distortions. The calculated DOS of MgRh$_2$O$_4$, Fig.~\ref{fig:dos-all} shows a filled valence band from -7.5 eV to -2.5 eV, dominantly of O-$p$ character, followed by Rh-$t_{2g}$ states from -2.5 to 0~eV. The bandgap originates from the crystal field splitting of the $d$ orbitals of about 1~eV. The conduction band is made of empty $e_g$ orbitals. This compound shows relatively small trigonal distortions compared to the other oxide spinels.

While the energy above \textit{hull} of 0.287 eV/atom for MgIr$_2$O$_4$ indicates instability of the spinel structure, the high stability of the MgRh$_2$O$_4$ spinel oxide has been calculated through the energy above \textit{hull} of 0.000 eV/atom. The calculated open-circuit voltages are 2.2~eV and 1.9~eV for the Rh and Ir spinels, respectively. Our calculations yield an increase of the open-circuit voltage for low Mg concentrations in both systems. Therefore, they are ideal candidates for cathode materials in Mg batteries with medium voltages.

Note that iridium and rhodium exhibit a low abundance in the earth's crust, making both iridium and rhodium to expensive raw materials. Battery materials should  be sustainable and associated with a high security of supply. However, there is always a compromise between gravimetric and volumetric energy density, abundance, cost and performance. Our calculations demonstrate that there are spinel oxides that promise a high Mg mobility as an electrode material. Possibly by doping less expensive oxide spinels with either iridium or rhodium a compromise between cost and performance can be achieved.

\subsection{Performance evaluation of the considered spinels}
After having separately discussed the considered spinel materials, we now summarize the results with respect to the ion mobility, and in the following we do the same with respect to the electrochemical and structural properties of the oxide spinels. Mn$_2$O$_4$ spinels have been successfully commercialized for use in Li-ion batteries. However, for this particular spinel, there is a high Mg$^{2+}$ diffusion barrier of 0.67~eV, which is higher than the Li$^+$ ion migration barriers in all M$_2$O$_4$ (M = Mn, Co, Ni, Cr) spinels which lie in the range 0.40-0.60~eV. Among the considered 3$d$ transition metals, Fe (0.94 eV) and V (0.78 eV) have the highest barriers for Mg$^{2+}$ ion diffusion, while Cr and Co are better conductors (0.51 eV). It should be noted that the migration barriers at low Mg concentrations are usually higher than those at high Mg concentrations. Thus, the Mg mobility becomes lower as the spinels change from the Mg-rich to the Mg-deficient limit.

Our calculations reveal that the tetrahedral sites are more stable for Mg than the octahedral sites for early transition metals in fully occupied Mg-spinels, while late transition metals such as Fe, Co, and Ni lead to a higher Mg stability in the octahedral sites. 

Going from the 3$d$ transition metals to the 4$d$ and 5$d$ metals such as Rh and Ir, the migration barriers obtained for Mg$^{2+}$ in these oxide spinels lie within 0.30-0.50~eV in the high Mg concentration limit. Accordingly, whereas Li$^+$ intercalates easily into various spinel oxides, Mg$^{2+}$ intercalation can benefit from more electronegative transition metals. Furthermore, these elements suppress the structural distortions coming from the antiferromagnetic order of the 3$d$-orbitals. We also propose that investigating doped systems with two different transition metals might lead to the idenfication of further oxide spinel materials with favorable properties as cathode materials for Mg batteries.

An additional essential parameter for the performance of a battery system is the open-circuit voltage. After understanding the electronic and structural properties, we address the principles that determine the crucial properties of electrochemical cells, voltages, and capacities.

The open-circuit voltage, $V_{OC}$, is obtained through the Mg intercalation energy $E_{inter}$ with respect to a metallic magnesium anode. In the Mg intercalation energy, the Gibbs free energy $G$ is approximated by the total 
energy $E$, neglecting the influence of zero-point energies and entropic contributions at finite temperatures.

\begin{figure}[!htb]
\centering
\includegraphics[width=0.60\linewidth]{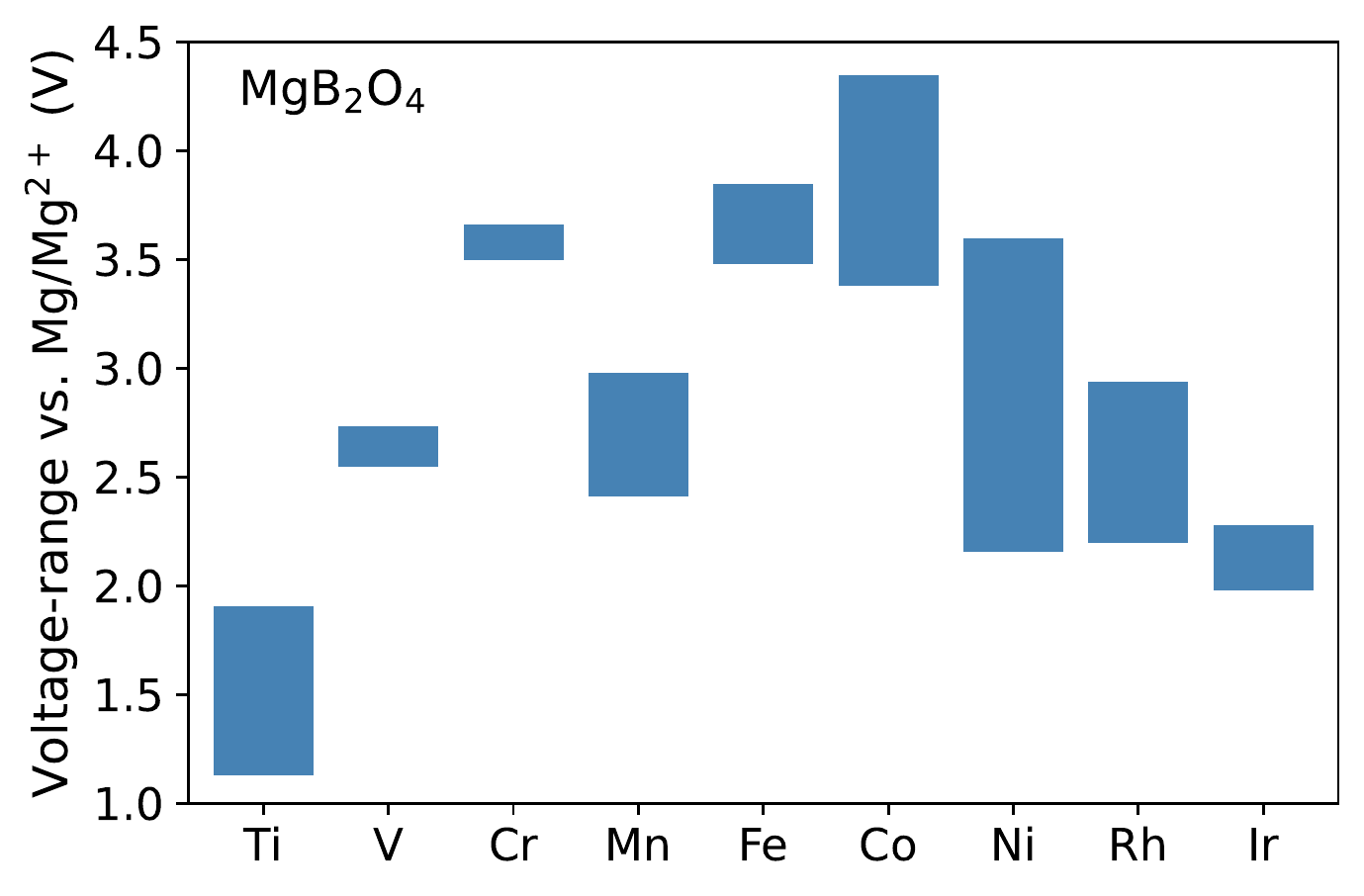}
\caption{\label{fig:voltage}The calculated voltage range for oxide spinel compounds as a function of the redox-active transition metals. The voltage range is obtained as the range between the calculated voltages for the low and high Mg concentrations of the considered compounds. }
\end{figure}

The calculated open-circuit voltages in the high magnesium concentrations indicate that the spinel oxides based on Cr, Mn, Fe, Co, and Ni exhibit high theoretical voltages for Mg intercalation. It should be noted that the voltages are referenced to the bulk metal phase of Mg metal. The Mg-based spinel oxides exhibit lower potentials for Mg intercalation (less than 4~V) compared with Li spinels that operate at potentials for Li intercation voltages 4 to 5~V. Except for Ti which has an Mg intercalation voltage lower than 2~V, most Mg intercalation potentials range between 2 and 4~V. However, the additional charge of the bivalent Mg$^{2+}$ causes a higher energy density in spinel cathodes compared to the monovalent Li$^{+}$. Hence, the considered Mg oxide spinels are viable 
candidates from the point of view of theoretical capacity.

The maximum theoretical specific energy (MTSE) corresponding to the energy per unit weight is given by
\begin{eqnarray}
\mathrm{MTSE} = (xV/W_t)F
\end{eqnarray}
where $W_t$ is the sum of the molecular weights of the reactants engaged in the intercalating reaction with a Mg concentration $x$, and $V$ stands for the average voltage of this reaction. $F$ is the Faraday constant.

Theoretically, most spinel oxides exhibit an average voltage of 2-4~V and correspondingly a high MTSE. Therefore, oxide spinels are currently the most promising candidates for high-performance Mg battery cathodes. 

It should be mentioned that in these materials cooperative Jahn–Teller distortions are the cause of volume contraction/expansion. In Tab.~\ref{tab:strc}, we also listed the calculated volume change $\Delta V/V$ upon Mg removal. Most of these materials exhibit volume contraction and expansion after demagnesiation of less than 3.7~\% compared to the structures full of Mg$^{2+}$. In the commercial LiCoO$_2$ cathode for Li-ion batteries, the cell volume shrinks by about ~5\% after delithiation. \cite{Wang_2007} Therefore, the volume changes for magnesium intercalation are relatively low for oxide spinel cathodes,  indicating a good cathode functionality with respect to volume change. 

\subsection{Synthesizability of Ir$^{3+}$ and Rh$^{3+}$ substituted spinel compounds}
Our calculations have demonstrated that Mg$^{2+}$ spinel oxides including 4$d$ and 5$d$ transition metals have promising properties as cathode materials for Mg batteries, in particular also with regard to ion mobility. Nevertheless, the feasibility of synthesizing pure compounds as well as solid solutions under laboratory conditions needs to be guaranteed. Literature research confirms the synthesisability of Rh$^{3+}$ containing spinel oxides as pure end-member compounds of Mg spinels like MgRh$_2$O$_4$~\cite{capobianco1993thermal,NELL19974159} or in solid solutions like Co$_{0.2}$Zn$_{0.8}$Fe$_{2-x}$Rh$_x$O$_4$~\cite{BHOWMIK200127} or CuCr$_{2-x}$Rh$_x$O$_4$~\cite{PADMANABAN1990286}. Therefore, a prosperous incorporation of Rh$^{3+}$ in these compounds seems promising and should be subjected to further investigation to find prospective candidates for optimised ionic conductors in the field of Mg-oxide spinel compounds.

In contrast to Rh, the 5$d$ equivalent Iridium show a different behavior. Just few oxidic compounds with Ir$^{3+}$, like in layered K$_{0.75}$Na$_{0.25}$IrO$_2$~\cite{weber2017}, are reported, as most oxides show Ir in oxidation states $\geq$ 4+. Miao and Seshadri have addressed this problem explaining it by relativistic effects~\cite{Miao_2012}. These relativistic effects lead to a distinct stabilization in oxidation states between those typical for Rh and Ir oxides. In contrast to Rh, which is stable as sesquioxide, the iridates exhibit a higher stability in dioxides, or at lower chemical potentials a decomposition to Ir and O$_2$ appear.  

These results are confirmed by our lab experiments in which we investigated possible solid solutions in the MgCr$_{2-x}$M$_x$O$_4$ (M: Rh, Ir) system. Different attempts to include Ir$^{3+}$ in the Magnesiochromite lattice failed, leading to the formation of either IrO$_2$ or Ir as second phases.

\begin{figure*}[!htb]
\centering
\includegraphics[width=1.0\linewidth]{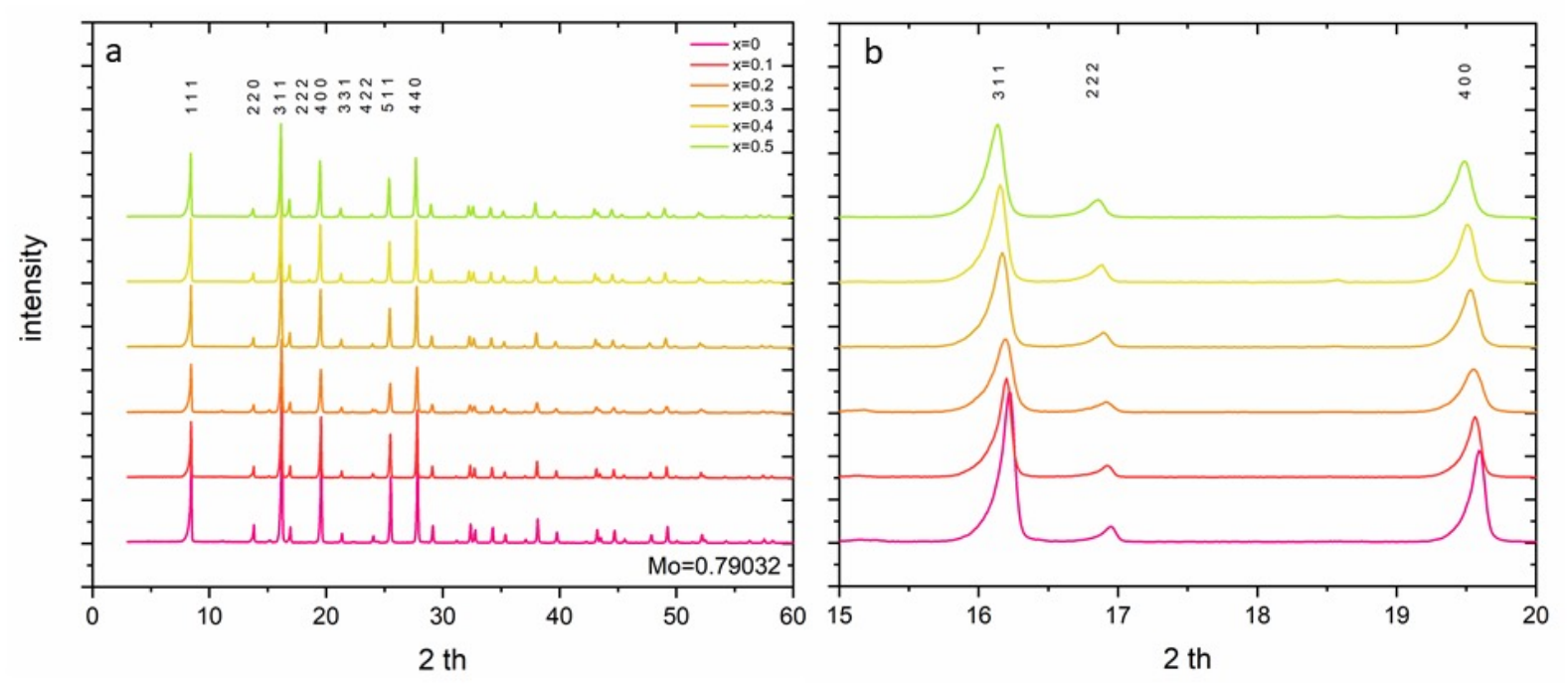}
\caption{\label{fig:xrd}XRD patterns at room temperature for MgCr$_{2-x}$Rh$_x$O$_4$ with x = 0-0.5 (a) full pattern; (b) section from 15-20$^\circ$ 2th showing the shift of peak positions.}
\end{figure*}

In contrast to that, initial results of our group suggest that the incorporation of Rh$^{3+}$ into the Magnesiochromite lattice is possible. Samples with different amounts of rhodium have been synthesized by solid state reaction of MgO (ABCR 99,5\%), Cr$_2$O$_3$ (Strem 99\%) and Rh$_2$O$_3$ (Santa Cruz Biotechnology). Stoichiometric amounts of the oxides were mixed in an agate mortar in acetone. Afterward the mixture was pressed into pellets and calcinated in alumina boats in a muffle oven (Nabertherm) for one week at 1000$^\circ$C with one intermediate grinding. The resulting brown powder was analysed by X-ray diffraction on a STOE StadiP diffractometer with monochromatic MoK$\alpha_1$ radiation. The XRD patterns of compounds with Rh content of x = 0-0.5 are depicted in Fig.~\ref{fig:xrd}a. An expected shift of the peak position can be clearly observed (Fig. ~\ref{fig:xrd}b) as well as a continuous change in the intensities. Rietveld refinements were performed with Fullprof~\cite{roisnel2001}. The results are given in Tab.~\ref{tab:crystal-info} where the crystallographic information is listed. 

\begin{figure}[!htb]
\centering
\includegraphics[width=0.50\linewidth]{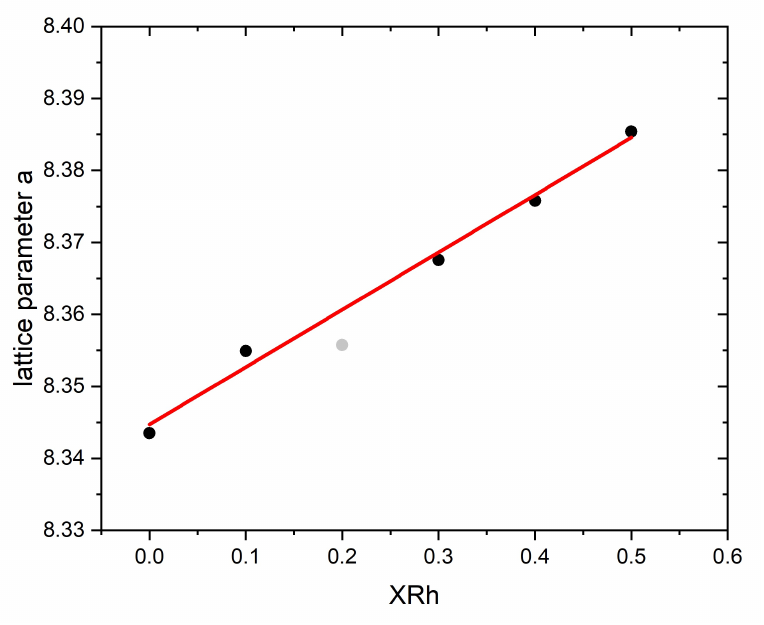}
\caption{\label{fig:fit}linear fit of lattice parameter with rising content of x$_{Rh}$ (grey: excluded sample); standard deviation is indistinguishable from symbols.}
\end{figure}

Beside the main phase, small amounts of 0.2 – 4.2\% of impurities are visible in the diffractograms. These consist of Cr$_{2-x}$Rh$_x$O$_3$ and Rh. The lattice parameters follow Vegard’s law, with a small deviation (see Fig.~\ref{fig:fit}). This sample shows the highest amount of impurity and differs by its color. In contrast to the others, this one is slightly green. The formation of a solid solution is therefore obvious, but must be investigated further, as well as the effects of the substitution on the physical properties. This will be part of a future work.

\begin{table*}[!htb]
\caption{\label{tab:crystal-info}Crystallographic information obtained by rietveld refinement.}
\begin{center}
\begin{tabular}{lccccc}
\hline
\hline
x$_{Rh}$  & Cell-parameter $a$ & Thermal displacement & Occupancy & Impurities & Chi$^2$ \\
by weight  & in {\AA} &  parameter $u$ in {\AA}$^2$ & Occ$_{Rh}$ & in wt\% &  \\
         \hline
0.0	& 8.34376	& 0.38601	& 0	            & -	                    				& 3.00 \\
0.1	& 8.35497	& 	0.38626 & 0.00416	& 1.0\% Cr$_2$O$_3$	& 1.70  \\
0.2	& 8.35589	& 0.38628	& 0.00248	& 4.2\% Cr$_2$O$_3$	& 2.25 \\
0.3	& 8.36755	& 0.38612	& 0.00539	& 0.2\% Rh	        			& 2.00 \\
0.4	& 8.37596	& 0.38605	& 0.01020	& 0.4\% Rh	        			& 1.51 \\
0.5	& 8.38555	& 0.38584	& 0.01316	& 0.3\% Rh      				& 1.80 \\
 \hline
 \hline
\end{tabular}
\end{center}
\end{table*} 

\section{Conclusions and Summary}
In this work, we have combined first-principles eletronic structure calculations and experimental preparation efforts to identify promising oxide spinel materials for Mg battery cathodes with superior Mg mobility properties. According to our calculations, oxide spinels with the electronegative $4d$ and $5d$ transition metals Rh and Ir allow such superior transport properties together with an acceptable open circuit voltage with respect to a Mg metal anode. The specific non-magnetic electronic structure of these oxide spinels make the compounds rather stable showing only relatively small trigonal distortions. Furthermore, we performed first attempts to synthesize these materials obtaining promising results, in particular with respect to Ir compounds. In general, our results and the analysis based on a combined experimental and theoretical study provide a conceptual framework to understand fast ion conductivity in oxide spinel electrode materials that will also be beneficial for the understanding and improvement of ion mobility in other materials classes.

\section*{Acknowledgements}
This work contributes to the research performed at CELEST (Center for 
Electrochemical Energy Storage Ulm-Karlsruhe) and was funded by the German 
Research Foundation (DFG) under Project ID 390874152 (POLiS Cluster of 
Excellence) and the Dr. Barbara Mez-Starck foundation. The authors acknowledge 
computer time provided by the state of Baden-W\"urttemberg through bwHPC and 
the German Research Foundation (DFG) through grant no INST 
40/575-1 FUGG (JUSTUS 2 cluster).


\begin{mcitethebibliography}{58}
\providecommand*\natexlab[1]{#1}
\providecommand*\mciteSetBstSublistMode[1]{}
\providecommand*\mciteSetBstMaxWidthForm[2]{}
\providecommand*\mciteBstWouldAddEndPuncttrue
  {\def\EndOfBibitem{\unskip.}}
\providecommand*\mciteBstWouldAddEndPunctfalse
  {\let\EndOfBibitem\relax}
\providecommand*\mciteSetBstMidEndSepPunct[3]{}
\providecommand*\mciteSetBstSublistLabelBeginEnd[3]{}
\providecommand*\EndOfBibitem{}
\mciteSetBstSublistMode{f}
\mciteSetBstMaxWidthForm{subitem}{(\alph{mcitesubitemcount})}
\mciteSetBstSublistLabelBeginEnd
  {\mcitemaxwidthsubitemform\space}
  {\relax}
  {\relax}

\bibitem[Hwang \latin{et~al.}(2017)Hwang, Myung, and Sun]{hwang17_CSR46_3529}
Hwang,~J.-Y.; Myung,~S.-T.; Sun,~Y.-K. Sodium-ion batteries: present and
  future. \emph{Chem. Soc. Rev.} \textbf{2017}, \emph{46}, 3529--3614\relax
\mciteBstWouldAddEndPuncttrue
\mciteSetBstMidEndSepPunct{\mcitedefaultmidpunct}
{\mcitedefaultendpunct}{\mcitedefaultseppunct}\relax
\EndOfBibitem
\bibitem[Yabuuchi \latin{et~al.}(2014)Yabuuchi, Kubota, Dahbi, and
  Komaba]{cnaoaki14_chemrev114_11636}
Yabuuchi,~N.; Kubota,~K.; Dahbi,~M.; Komaba,~S. Research Development on
  Sodium-Ion Batteries. \emph{Chem. Rev.} \textbf{2014}, \emph{114},
  11636--11682\relax
\mciteBstWouldAddEndPuncttrue
\mciteSetBstMidEndSepPunct{\mcitedefaultmidpunct}
{\mcitedefaultendpunct}{\mcitedefaultseppunct}\relax
\EndOfBibitem
\bibitem[Gregory \latin{et~al.}(1990)Gregory, Hoffman, and
  Winterton]{gregory90_jelso137_775}
Gregory,~T.~D.; Hoffman,~R.~J.; Winterton,~R.~C. Nonaqueous Electrochemistry of
  Magnesium: Applications to Energy Storage. \emph{J. Electrochem. Soc.}
  \textbf{1990}, \emph{137}, 775--780\relax
\mciteBstWouldAddEndPuncttrue
\mciteSetBstMidEndSepPunct{\mcitedefaultmidpunct}
{\mcitedefaultendpunct}{\mcitedefaultseppunct}\relax
\EndOfBibitem
\bibitem[Aurbach \latin{et~al.}(2000)Aurbach, Lu, Schechter, Gofer, Gizbar,
  Turgeman, Cohen, Moshkovich, and Levi]{aurbach00_nature407_724}
Aurbach,~D.; Lu,~Z.; Schechter,~A.; Gofer,~Y.; Gizbar,~H.; Turgeman,~R.;
  Cohen,~Y.; Moshkovich,~M.; Levi,~E. Prototype systems for rechargeable
  magnesium batteries. \emph{Nature} \textbf{2000}, \emph{407}, 724--727\relax
\mciteBstWouldAddEndPuncttrue
\mciteSetBstMidEndSepPunct{\mcitedefaultmidpunct}
{\mcitedefaultendpunct}{\mcitedefaultseppunct}\relax
\EndOfBibitem
\bibitem[MacLaughlin(2019)]{MacLaughlin2019}
MacLaughlin,~C.~M. Status and Outlook for Magnesium Battery Technologies: A
  Conversation with {Stan Whittingham} and {Sarbajit Banerjee}. \emph{ACS
  Energy Lett.} \textbf{2019}, \emph{4}, 572--575\relax
\mciteBstWouldAddEndPuncttrue
\mciteSetBstMidEndSepPunct{\mcitedefaultmidpunct}
{\mcitedefaultendpunct}{\mcitedefaultseppunct}\relax
\EndOfBibitem
\bibitem[Davidson \latin{et~al.}(2020)Davidson, Verma, Santos, Hao, Fincher,
  Zhao, Attari, Schofield, Van~Buskirk, Fraticelli-Cartagena, Alivio, Arroyave,
  Xie, Pharr, Mukherjee, and Banerjee]{Davidson2020}
Davidson,~R. \latin{et~al.}  Mapping mechanisms and growth regimes of magnesium
  electrodeposition at high current densities. \emph{Mater. Horiz.}
  \textbf{2020}, \emph{7}, 843--854\relax
\mciteBstWouldAddEndPuncttrue
\mciteSetBstMidEndSepPunct{\mcitedefaultmidpunct}
{\mcitedefaultendpunct}{\mcitedefaultseppunct}\relax
\EndOfBibitem
\bibitem[Singh \latin{et~al.}(2013)Singh, Arthur, Ling, Matsui, and
  Mizuno]{singh13_chemcom49_149}
Singh,~N.; Arthur,~T.~S.; Ling,~C.; Matsui,~M.; Mizuno,~F. A high
  energy-density tin anode for rechargeable magnesium-ion batteries.
  \emph{Chem. Commun.} \textbf{2013}, \emph{49}, 149--151\relax
\mciteBstWouldAddEndPuncttrue
\mciteSetBstMidEndSepPunct{\mcitedefaultmidpunct}
{\mcitedefaultendpunct}{\mcitedefaultseppunct}\relax
\EndOfBibitem
\bibitem[Zhao-Karger \latin{et~al.}(2017)Zhao-Karger, Gil~Bardaji, Fuhr, and
  Fichtner]{zhao-karger17_jmatchema5_10815}
Zhao-Karger,~Z.; Gil~Bardaji,~M.~E.; Fuhr,~O.; Fichtner,~M. A new class of
  non-corrosive{,} highly efficient electrolytes for rechargeable magnesium
  batteries. \emph{J. Mater. Chem. A} \textbf{2017}, \emph{5},
  10815--10820\relax
\mciteBstWouldAddEndPuncttrue
\mciteSetBstMidEndSepPunct{\mcitedefaultmidpunct}
{\mcitedefaultendpunct}{\mcitedefaultseppunct}\relax
\EndOfBibitem
\bibitem[Sun \latin{et~al.}(2016)Sun, Bonnick, Duffort, Liu, Rong, Persson,
  Ceder, and Nazar]{sun16_energyensci9_2273}
Sun,~X.; Bonnick,~P.; Duffort,~V.; Liu,~M.; Rong,~Z.; Persson,~K.~A.;
  Ceder,~G.; Nazar,~L.~F. A high capacity thiospinel cathode for {Mg}
  batteries. \emph{Energy Environ. Sci.} \textbf{2016}, \emph{9},
  2273--2277\relax
\mciteBstWouldAddEndPuncttrue
\mciteSetBstMidEndSepPunct{\mcitedefaultmidpunct}
{\mcitedefaultendpunct}{\mcitedefaultseppunct}\relax
\EndOfBibitem
\bibitem[Canepa \latin{et~al.}(2017)Canepa, Sai~Gautam, Hannah, Malik, Liu,
  Gallagher, Persson, and Ceder]{canepa17_chemrev117_4287}
Canepa,~P.; Sai~Gautam,~G.; Hannah,~D.~C.; Malik,~R.; Liu,~M.;
  Gallagher,~K.~G.; Persson,~K.~A.; Ceder,~G. Odyssey of Multivalent Cathode
  Materials: Open Questions and Future Challenges. \emph{Chem. Rev.}
  \textbf{2017}, \emph{117}, 4287--4341\relax
\mciteBstWouldAddEndPuncttrue
\mciteSetBstMidEndSepPunct{\mcitedefaultmidpunct}
{\mcitedefaultendpunct}{\mcitedefaultseppunct}\relax
\EndOfBibitem
\bibitem[Orikasa \latin{et~al.}(2014)Orikasa, Masese, Koyama, Mori, Hattori,
  Yamamoto, Okado, Huang, Minato, Tassel, Kim, Kobayashi, Abe, Kageyama, and
  Uchimoto]{orikasa14_sr4_5622}
Orikasa,~Y.; Masese,~T.; Koyama,~Y.; Mori,~T.; Hattori,~M.; Yamamoto,~K.;
  Okado,~T.; Huang,~Z.-D.; Minato,~T.; Tassel,~C.; Kim,~J.; Kobayashi,~Y.;
  Abe,~T.; Kageyama,~H.; Uchimoto,~Y. High energy density rechargeable
  magnesium battery using earth-abundant and non-toxic elements. \emph{Sci.
  Rep.} \textbf{2014}, \emph{4}, 5622\relax
\mciteBstWouldAddEndPuncttrue
\mciteSetBstMidEndSepPunct{\mcitedefaultmidpunct}
{\mcitedefaultendpunct}{\mcitedefaultseppunct}\relax
\EndOfBibitem
\bibitem[NuLi \latin{et~al.}(2009)NuLi, Yang, Wang, and
  Li]{nuli09_jpcc113_12594}
NuLi,~Y.; Yang,~J.; Wang,~J.; Li,~Y. Electrochemical Intercalation of Mg$^{2+}$
  in Magnesium Manganese Silicate and Its Application as High-Energy
  Rechargeable Magnesium Battery Cathode. \emph{J. Phys. Chem. C}
  \textbf{2009}, \emph{113}, 12594--12597\relax
\mciteBstWouldAddEndPuncttrue
\mciteSetBstMidEndSepPunct{\mcitedefaultmidpunct}
{\mcitedefaultendpunct}{\mcitedefaultseppunct}\relax
\EndOfBibitem
\bibitem[Ling and Mizuno(2013)Ling, and Mizuno]{ling13_cm25_3062}
Ling,~C.; Mizuno,~F. Phase Stability of Post-spinel Compound AMn2O4 (A = Li,
  Na, or Mg) and Its Application as a Rechargeable Battery Cathode. \emph{Chem.
  Mater.} \textbf{2013}, \emph{25}, 3062--3071\relax
\mciteBstWouldAddEndPuncttrue
\mciteSetBstMidEndSepPunct{\mcitedefaultmidpunct}
{\mcitedefaultendpunct}{\mcitedefaultseppunct}\relax
\EndOfBibitem
\bibitem[Liu \latin{et~al.}(2015)Liu, Rong, Malik, Canepa, Jain, Ceder, and
  Persson]{liu15_energyensci8_964}
Liu,~M.; Rong,~Z.; Malik,~R.; Canepa,~P.; Jain,~A.; Ceder,~G.; Persson,~K.~A.
  Spinel compounds as multivalent battery cathodes: a systematic evaluation
  based on ab initio calculations. \emph{Energy Environ. Sci.} \textbf{2015},
  \emph{8}, 964--974\relax
\mciteBstWouldAddEndPuncttrue
\mciteSetBstMidEndSepPunct{\mcitedefaultmidpunct}
{\mcitedefaultendpunct}{\mcitedefaultseppunct}\relax
\EndOfBibitem
\bibitem[Ling \latin{et~al.}(2015)Ling, Zhang, Arthur, and
  Mizuno]{ling15_cm27_5799}
Ling,~C.; Zhang,~R.; Arthur,~T.~S.; Mizuno,~F. How General is the Conversion
  Reaction in Mg Battery Cathode: A Case Study of the Magnesiation of
  $\alpha$-MnO2. \emph{Chem. Mater.} \textbf{2015}, \emph{27}, 5799--5807\relax
\mciteBstWouldAddEndPuncttrue
\mciteSetBstMidEndSepPunct{\mcitedefaultmidpunct}
{\mcitedefaultendpunct}{\mcitedefaultseppunct}\relax
\EndOfBibitem
\bibitem[Kim \latin{et~al.}(2015)Kim, Phillips, Key, Yi, Nordlund, Yu, Bayliss,
  Han, He, Zhang, Burrell, Klie, and Cabana]{kim15_advmat27_3377}
Kim,~C.; Phillips,~P.~J.; Key,~B.; Yi,~T.; Nordlund,~D.; Yu,~Y.-S.;
  Bayliss,~R.~D.; Han,~S.-D.; He,~M.; Zhang,~Z.; Burrell,~A.~K.; Klie,~R.~F.;
  Cabana,~J. Direct Observation of Reversible Magnesium Ion Intercalation into
  a Spinel Oxide Host. \emph{Adv. Mater.} \textbf{2015}, \emph{27},
  3377--3384\relax
\mciteBstWouldAddEndPuncttrue
\mciteSetBstMidEndSepPunct{\mcitedefaultmidpunct}
{\mcitedefaultendpunct}{\mcitedefaultseppunct}\relax
\EndOfBibitem
\bibitem[Dillenz \latin{et~al.}(2020)Dillenz, Sotoudeh, Euchner, and
  Gro{\ss}]{dillenz20_fer8_260}
Dillenz,~M.; Sotoudeh,~M.; Euchner,~H.; Gro{\ss},~A. Screening of Charge
  Carrier Migration in the {MgSc$_2$Se$_4$} Spinel Structure. \emph{Front.
  Energy Res.} \textbf{2020}, \emph{8}, 260\relax
\mciteBstWouldAddEndPuncttrue
\mciteSetBstMidEndSepPunct{\mcitedefaultmidpunct}
{\mcitedefaultendpunct}{\mcitedefaultseppunct}\relax
\EndOfBibitem
\bibitem[Sotoudeh \latin{et~al.}(2021)Sotoudeh, Dillenz, and
  Gro{\ss}]{Sotoudeh_Mg_Spinel}
Sotoudeh,~M.; Dillenz,~M.; Gro{\ss},~A. Mechanism of Magnesium Transport in
  Spinel Chalcogenides. \emph{Adv. Energy Sustainability Res.} \textbf{2021},
  \emph{2}, 2100113\relax
\mciteBstWouldAddEndPuncttrue
\mciteSetBstMidEndSepPunct{\mcitedefaultmidpunct}
{\mcitedefaultendpunct}{\mcitedefaultseppunct}\relax
\EndOfBibitem
\bibitem[Sotoudeh and Gro{\ss}(2022)Sotoudeh, and
  Gro{\ss}]{Sotoudeh_Descriptor}
Sotoudeh,~M.; Gro{\ss},~A. Descriptor and Scaling Relations for Ion Mobility in
  Crystalline Solids. \emph{JACS Au} \textbf{2022}, \emph{2}, 463--471\relax
\mciteBstWouldAddEndPuncttrue
\mciteSetBstMidEndSepPunct{\mcitedefaultmidpunct}
{\mcitedefaultendpunct}{\mcitedefaultseppunct}\relax
\EndOfBibitem
\bibitem[Sotoudeh and Gro{\ss}(2022)Sotoudeh, and
  Gro{\ss}]{sotoudeh22_jpcl13_10092}
Sotoudeh,~M.; Gro{\ss},~A. Stability of Magnesium Binary and Ternary Compounds
  for Batteries Determined from First Principles. \emph{J. Phys. Chem. Lett.}
  \textbf{2022}, \emph{13}, 10092--10100\relax
\mciteBstWouldAddEndPuncttrue
\mciteSetBstMidEndSepPunct{\mcitedefaultmidpunct}
{\mcitedefaultendpunct}{\mcitedefaultseppunct}\relax
\EndOfBibitem
\bibitem[Bayliss \latin{et~al.}(2020)Bayliss, Key, Sai~Gautam, Canepa, Kwon,
  Lapidus, Dogan, Adil, Lipton, Baker, Ceder, Vaughey, and
  Cabana]{Bayliss20_chemmat32_663}
Bayliss,~R.~D.; Key,~B.; Sai~Gautam,~G.; Canepa,~P.; Kwon,~B.~J.;
  Lapidus,~S.~H.; Dogan,~F.; Adil,~A.~A.; Lipton,~A.~S.; Baker,~P.~J.;
  Ceder,~G.; Vaughey,~J.~T.; Cabana,~J. Probing Mg Migration in Spinel Oxides.
  \emph{Chem. Mater.} \textbf{2020}, \emph{32}, 663--670\relax
\mciteBstWouldAddEndPuncttrue
\mciteSetBstMidEndSepPunct{\mcitedefaultmidpunct}
{\mcitedefaultendpunct}{\mcitedefaultseppunct}\relax
\EndOfBibitem
\bibitem[Bachman \latin{et~al.}(2016)Bachman, Muy, Grimaud, Chang, Pour, Lux,
  Paschos, Maglia, Lupart, Lamp, Giordano, and
  Shao-Horn]{bachman16_chemrev116_140}
Bachman,~J.~C.; Muy,~S.; Grimaud,~A.; Chang,~H.-H.; Pour,~N.; Lux,~S.~F.;
  Paschos,~O.; Maglia,~F.; Lupart,~S.; Lamp,~P.; Giordano,~L.; Shao-Horn,~Y.
  Inorganic Solid-State Electrolytes for Lithium Batteries: Mechanisms and
  Properties Governing Ion Conduction. \emph{Chem. Rev.} \textbf{2016},
  \emph{116}, 140--162\relax
\mciteBstWouldAddEndPuncttrue
\mciteSetBstMidEndSepPunct{\mcitedefaultmidpunct}
{\mcitedefaultendpunct}{\mcitedefaultseppunct}\relax
\EndOfBibitem
\bibitem[Yabuuchi \latin{et~al.}(2015)Yabuuchi, Takeuchi, Nakayama, Shiiba,
  Ogawa, Nakayama, Ohta, Endo, Ozaki, Inamasu, Sato, and Komaba]{Yabuuchi7650}
Yabuuchi,~N.; Takeuchi,~M.; Nakayama,~M.; Shiiba,~H.; Ogawa,~M.; Nakayama,~K.;
  Ohta,~T.; Endo,~D.; Ozaki,~T.; Inamasu,~T.; Sato,~K.; Komaba,~S.
  High-capacity electrode materials for rechargeable lithium batteries:
  Li3NbO4-based system with cation-disordered rocksalt structure. \emph{Proc.
  Natl. Acad. Sci. U.S.A.} \textbf{2015}, \emph{112}, 7650--7655\relax
\mciteBstWouldAddEndPuncttrue
\mciteSetBstMidEndSepPunct{\mcitedefaultmidpunct}
{\mcitedefaultendpunct}{\mcitedefaultseppunct}\relax
\EndOfBibitem
\bibitem[Koettgen \latin{et~al.}(2020)Koettgen, Bartel, and
  Ceder]{Koettgen2020}
Koettgen,~J.; Bartel,~C.~J.; Ceder,~G. Computational investigation of
  chalcogenide spinel conductors for all-solid-state {Mg} batteries.
  \emph{Chem. Commun.} \textbf{2020}, \emph{56}, 1952--1955\relax
\mciteBstWouldAddEndPuncttrue
\mciteSetBstMidEndSepPunct{\mcitedefaultmidpunct}
{\mcitedefaultendpunct}{\mcitedefaultseppunct}\relax
\EndOfBibitem
\bibitem[Ben~Yahia \latin{et~al.}(2019)Ben~Yahia, Vergnet, Sauban{\`e}re, and
  Doublet]{Yahia2019}
Ben~Yahia,~M.; Vergnet,~J.; Sauban{\`e}re,~M.; Doublet,~M.-L. Unified picture
  of anionic redox in {Li/Na}-ion batteries. \emph{Nat. Mater.} \textbf{2019},
  \emph{18}, 496--502\relax
\mciteBstWouldAddEndPuncttrue
\mciteSetBstMidEndSepPunct{\mcitedefaultmidpunct}
{\mcitedefaultendpunct}{\mcitedefaultseppunct}\relax
\EndOfBibitem
\bibitem[Hohenberg and Kohn(1964)Hohenberg, and Kohn]{hohenberg64_pr136_B864}
Hohenberg,~P.; Kohn,~W. Inhomogeneous Electron Gas. \emph{Phys. Rev.}
  \textbf{1964}, \emph{136}, B864--B871\relax
\mciteBstWouldAddEndPuncttrue
\mciteSetBstMidEndSepPunct{\mcitedefaultmidpunct}
{\mcitedefaultendpunct}{\mcitedefaultseppunct}\relax
\EndOfBibitem
\bibitem[Kohn and Sham(1965)Kohn, and Sham]{kohn65_pr140_1133}
Kohn,~W.; Sham,~L.~J. Self-Consistent Equations Including Exchange and
  Correlation Effects. \emph{Phys. Rev.} \textbf{1965}, \emph{140},
  A1133--A1138\relax
\mciteBstWouldAddEndPuncttrue
\mciteSetBstMidEndSepPunct{\mcitedefaultmidpunct}
{\mcitedefaultendpunct}{\mcitedefaultseppunct}\relax
\EndOfBibitem
\bibitem[Euchner and Gro{\ss}(2022)Euchner, and Gro{\ss}]{Euchner2022}
Euchner,~H.; Gro{\ss},~A. Atomistic modeling of {Li}- and post-{Li-}ion
  batteries. \emph{Phys. Rev. Materials} \textbf{2022}, \emph{6}, 040302\relax
\mciteBstWouldAddEndPuncttrue
\mciteSetBstMidEndSepPunct{\mcitedefaultmidpunct}
{\mcitedefaultendpunct}{\mcitedefaultseppunct}\relax
\EndOfBibitem
\bibitem[Perdew \latin{et~al.}(1996)Perdew, Burke, and
  Ernzerhof]{perdew96_prl77_3865}
Perdew,~J.~P.; Burke,~K.; Ernzerhof,~M. Generalized Gradient Approximation Made
  Simple. \emph{Phys. Rev. Lett.} \textbf{1996}, \emph{77}, 3865--3868\relax
\mciteBstWouldAddEndPuncttrue
\mciteSetBstMidEndSepPunct{\mcitedefaultmidpunct}
{\mcitedefaultendpunct}{\mcitedefaultseppunct}\relax
\EndOfBibitem
\bibitem[Bl\"ochl(1994)]{bloechl94_prb50_17953}
Bl\"ochl,~P.~E. Projector augmented-wave method. \emph{Phys. Rev. B}
  \textbf{1994}, \emph{50}, 17953--17979\relax
\mciteBstWouldAddEndPuncttrue
\mciteSetBstMidEndSepPunct{\mcitedefaultmidpunct}
{\mcitedefaultendpunct}{\mcitedefaultseppunct}\relax
\EndOfBibitem
\bibitem[Kresse and Hafner(1993)Kresse, and Hafner]{kresse93_prb47_558}
Kresse,~G.; Hafner,~J. Ab initio molecular dynamics for liquid metals.
  \emph{Phys. Rev. B} \textbf{1993}, \emph{47}, 558--561\relax
\mciteBstWouldAddEndPuncttrue
\mciteSetBstMidEndSepPunct{\mcitedefaultmidpunct}
{\mcitedefaultendpunct}{\mcitedefaultseppunct}\relax
\EndOfBibitem
\bibitem[Kresse and Furthm\"uller(1996)Kresse, and
  Furthm\"uller]{kresse96_prb54_11169}
Kresse,~G.; Furthm\"uller,~J. Efficient iterative schemes for ab initio
  total-energy calculations using a plane-wave basis set. \emph{Phys. Rev. B}
  \textbf{1996}, \emph{54}, 11169--11186\relax
\mciteBstWouldAddEndPuncttrue
\mciteSetBstMidEndSepPunct{\mcitedefaultmidpunct}
{\mcitedefaultendpunct}{\mcitedefaultseppunct}\relax
\EndOfBibitem
\bibitem[Kresse and Joubert(1999)Kresse, and Joubert]{kresse99_prb59_1758}
Kresse,~G.; Joubert,~D. From ultrasoft pseudopotentials to the projector
  augmented-wave method. \emph{Phys. Rev. B} \textbf{1999}, \emph{59},
  1758--1775\relax
\mciteBstWouldAddEndPuncttrue
\mciteSetBstMidEndSepPunct{\mcitedefaultmidpunct}
{\mcitedefaultendpunct}{\mcitedefaultseppunct}\relax
\EndOfBibitem
\bibitem[Jepson and Anderson(1971)Jepson, and Anderson]{jepsen71_ssc9_1763}
Jepson,~O.; Anderson,~O.~K. The electronic structure of h.c.p. Ytterbium.
  \emph{Solid State Commun.} \textbf{1971}, \emph{9}, 1763--1767\relax
\mciteBstWouldAddEndPuncttrue
\mciteSetBstMidEndSepPunct{\mcitedefaultmidpunct}
{\mcitedefaultendpunct}{\mcitedefaultseppunct}\relax
\EndOfBibitem
\bibitem[Lehmann and Taut(1972)Lehmann, and Taut]{lehmann72_pssb54_469}
Lehmann,~G.; Taut,~M. On the Numerical Calculation of the Density of States and
  Related Properties. \emph{Phys. Stat. Sol. (b)} \textbf{1972}, \emph{54},
  469\relax
\mciteBstWouldAddEndPuncttrue
\mciteSetBstMidEndSepPunct{\mcitedefaultmidpunct}
{\mcitedefaultendpunct}{\mcitedefaultseppunct}\relax
\EndOfBibitem
\bibitem[Bl\"ochl \latin{et~al.}(1994)Bl\"ochl, Jepsen, and
  Andersen]{bloechl94_prb49_16223}
Bl\"ochl,~P.~E.; Jepsen,~O.; Andersen,~O.~K. Improved tetrahedron method for
  Brillouin-zone integrations. \emph{Phys. Rev. B} \textbf{1994}, \emph{49},
  16223--16233\relax
\mciteBstWouldAddEndPuncttrue
\mciteSetBstMidEndSepPunct{\mcitedefaultmidpunct}
{\mcitedefaultendpunct}{\mcitedefaultseppunct}\relax
\EndOfBibitem
\bibitem[Dudarev \latin{et~al.}(1998)Dudarev, Botton, Savrasov, Humphreys, and
  Sutton]{Dudarev98_prb57_1505}
Dudarev,~S.~L.; Botton,~G.~A.; Savrasov,~S.~Y.; Humphreys,~C.~J.; Sutton,~A.~P.
  Electron-energy-loss spectra and the structural stability of nickel oxide: An
  LSDA+U study. \emph{Phys. Rev. B} \textbf{1998}, \emph{57}, 1505--1509\relax
\mciteBstWouldAddEndPuncttrue
\mciteSetBstMidEndSepPunct{\mcitedefaultmidpunct}
{\mcitedefaultendpunct}{\mcitedefaultseppunct}\relax
\EndOfBibitem
\bibitem[Sheppard \latin{et~al.}(2008)Sheppard, Terrell, and
  Henkelman]{sheppard08_jcp128_134106}
Sheppard,~D.; Terrell,~R.; Henkelman,~G. Optimization methods for finding
  minimum energy paths. \emph{J. Chem. Phys.} \textbf{2008}, \emph{128},
  134106\relax
\mciteBstWouldAddEndPuncttrue
\mciteSetBstMidEndSepPunct{\mcitedefaultmidpunct}
{\mcitedefaultendpunct}{\mcitedefaultseppunct}\relax
\EndOfBibitem
\bibitem[Mehrer(2007)]{mehrer2007diffusion}
Mehrer,~H. \emph{Diffusion in Solids: Fundamentals, Methods, Materials,
  Diffusion-Controlled Processes}; Springer Series in Solid-State Sciences;
  Springer Berlin Heidelberg, 2007\relax
\mciteBstWouldAddEndPuncttrue
\mciteSetBstMidEndSepPunct{\mcitedefaultmidpunct}
{\mcitedefaultendpunct}{\mcitedefaultseppunct}\relax
\EndOfBibitem
\bibitem[Heyd \latin{et~al.}(2003)Heyd, Scuseria, and Ernzerhof]{HSE06}
Heyd,~J.; Scuseria,~G.~E.; Ernzerhof,~M. Hybrid functionals based on a screened
  Coulomb potential. \emph{J. Chem. Phys.} \textbf{2003}, \emph{118},
  8207--8215\relax
\mciteBstWouldAddEndPuncttrue
\mciteSetBstMidEndSepPunct{\mcitedefaultmidpunct}
{\mcitedefaultendpunct}{\mcitedefaultseppunct}\relax
\EndOfBibitem
\bibitem[Bartel(2022)]{Bartel22_jmc57_10475}
Bartel,~C.~J. Review of computational approaches to predict the thermodynamic
  stability of inorganic solids. \emph{J. Mater. Sci.} \textbf{2022},
  \emph{57}, 10475--10498\relax
\mciteBstWouldAddEndPuncttrue
\mciteSetBstMidEndSepPunct{\mcitedefaultmidpunct}
{\mcitedefaultendpunct}{\mcitedefaultseppunct}\relax
\EndOfBibitem
\bibitem[Jain \latin{et~al.}(2013)Jain, Ong, Hautier, Chen, Richards, Dacek,
  Cholia, Gunter, Skinner, Ceder, and Persson]{Jain13_apl1_011002}
Jain,~A.; Ong,~S.~P.; Hautier,~G.; Chen,~W.; Richards,~W.~D.; Dacek,~S.;
  Cholia,~S.; Gunter,~D.; Skinner,~D.; Ceder,~G.; Persson,~K.~A. Commentary:
  The Materials Project: A materials genome approach to accelerating materials
  innovation. \emph{APL Mater.} \textbf{2013}, \emph{1}, 011002\relax
\mciteBstWouldAddEndPuncttrue
\mciteSetBstMidEndSepPunct{\mcitedefaultmidpunct}
{\mcitedefaultendpunct}{\mcitedefaultseppunct}\relax
\EndOfBibitem
\bibitem[Wang \latin{et~al.}(2006)Wang, Maxisch, and
  Ceder]{Wang06_prb73_195107}
Wang,~L.; Maxisch,~T.; Ceder,~G. Oxidation energies of transition metal oxides
  within the $\mathrm{GGA}+\mathrm{U}$ framework. \emph{Phys. Rev. B}
  \textbf{2006}, \emph{73}, 195107\relax
\mciteBstWouldAddEndPuncttrue
\mciteSetBstMidEndSepPunct{\mcitedefaultmidpunct}
{\mcitedefaultendpunct}{\mcitedefaultseppunct}\relax
\EndOfBibitem
\bibitem[Jain \latin{et~al.}(2011)Jain, Hautier, Ong, Moore, Fischer, Persson,
  and Ceder]{Jain11_prb84_045115}
Jain,~A.; Hautier,~G.; Ong,~S.~P.; Moore,~C.~J.; Fischer,~C.~C.;
  Persson,~K.~A.; Ceder,~G. Formation enthalpies by mixing GGA and GGA $+$ $U$
  calculations. \emph{Phys. Rev. B} \textbf{2011}, \emph{84}, 045115\relax
\mciteBstWouldAddEndPuncttrue
\mciteSetBstMidEndSepPunct{\mcitedefaultmidpunct}
{\mcitedefaultendpunct}{\mcitedefaultseppunct}\relax
\EndOfBibitem
\bibitem[Burdett \latin{et~al.}(1982)Burdett, Price, and Price]{Burdett1982}
Burdett,~J.~K.; Price,~G.~D.; Price,~S.~L. Role of the crystal-field theory in
  determining the structures of spinels. \emph{J. Am. Chem. Soc.}
  \textbf{1982}, \emph{104}, 92--95\relax
\mciteBstWouldAddEndPuncttrue
\mciteSetBstMidEndSepPunct{\mcitedefaultmidpunct}
{\mcitedefaultendpunct}{\mcitedefaultseppunct}\relax
\EndOfBibitem
\bibitem[Banerjee \latin{et~al.}(1967)Banerjee, O'Reilly, Gibb, and
  Greenwood]{Banerjee1967}
Banerjee,~S.~K.; O'Reilly,~W.; Gibb,~T.; Greenwood,~N. The behaviour of ferrous
  ions in iron-titanium spinels. \emph{J. Phys. Chem. Sol.} \textbf{1967},
  \emph{28}, 1323 -- 1335\relax
\mciteBstWouldAddEndPuncttrue
\mciteSetBstMidEndSepPunct{\mcitedefaultmidpunct}
{\mcitedefaultendpunct}{\mcitedefaultseppunct}\relax
\EndOfBibitem
\bibitem[Sickafus \latin{et~al.}(1999)Sickafus, Wills, and Grimes]{Sickafus99}
Sickafus,~K.~E.; Wills,~J.~M.; Grimes,~N.~W. Structure of Spinel. \emph{J. Am.
  Ceram. Soc.} \textbf{1999}, \emph{82}, 3279--3292\relax
\mciteBstWouldAddEndPuncttrue
\mciteSetBstMidEndSepPunct{\mcitedefaultmidpunct}
{\mcitedefaultendpunct}{\mcitedefaultseppunct}\relax
\EndOfBibitem
\bibitem[Shannon(1976)]{Shannon1976}
Shannon,~R.~D. {Revised effective ionic radii and systematic studies of
  interatomic distances in halides and chalcogenides}. \emph{Acta Cryst. A}
  \textbf{1976}, \emph{32}, 751--767\relax
\mciteBstWouldAddEndPuncttrue
\mciteSetBstMidEndSepPunct{\mcitedefaultmidpunct}
{\mcitedefaultendpunct}{\mcitedefaultseppunct}\relax
\EndOfBibitem
\bibitem[Lacroix \latin{et~al.}(2011)Lacroix, Mendels, and
  Mila]{lacroix2011magnetism}
Lacroix,~C.; Mendels,~P.; Mila,~F. \emph{Introduction to Frustrated Magnetism:
  Materials, Experiments, Theory}; Springer Series in Solid-State Sciences;
  Springer Berlin Heidelberg, 2011\relax
\mciteBstWouldAddEndPuncttrue
\mciteSetBstMidEndSepPunct{\mcitedefaultmidpunct}
{\mcitedefaultendpunct}{\mcitedefaultseppunct}\relax
\EndOfBibitem
\bibitem[Wang \latin{et~al.}(2007)Wang, Sone, Segami, Naito, Yamada, and
  Kibe]{Wang_2007}
Wang,~X.; Sone,~Y.; Segami,~G.; Naito,~H.; Yamada,~C.; Kibe,~K. Understanding
  Volume Change in Lithium-Ion Cells during Charging and Discharging Using In
  Situ Measurements. \emph{J. Electrochem. Soc.} \textbf{2007}, \emph{154},
  A14\relax
\mciteBstWouldAddEndPuncttrue
\mciteSetBstMidEndSepPunct{\mcitedefaultmidpunct}
{\mcitedefaultendpunct}{\mcitedefaultseppunct}\relax
\EndOfBibitem
\bibitem[Capobianco(1993)]{capobianco1993thermal}
Capobianco,~C.~J. On the thermal decomposition of MgRh$_2$O$_4$ spinel and the
  solid solution Mg(Rh, Al)$_2$O$_4$. \emph{Thermochim. Acta} \textbf{1993},
  \emph{220}, 7--16\relax
\mciteBstWouldAddEndPuncttrue
\mciteSetBstMidEndSepPunct{\mcitedefaultmidpunct}
{\mcitedefaultendpunct}{\mcitedefaultseppunct}\relax
\EndOfBibitem
\bibitem[Nell and O'Neill(1997)Nell, and O'Neill]{NELL19974159}
Nell,~J.; O'Neill,~H.~S. The Gibbs free energy of formation and heat capacity
  of $\beta$-Rh$_2$O$_3$ and MgRh$_2$O$_4$, the MgO-Rh-O phase diagram, and
  constraints on the stability of Mg$_2$Rh$^{4+}$O$_4$. \emph{Geochim.
  Cosmochim. Acta} \textbf{1997}, \emph{61}, 4159--4171\relax
\mciteBstWouldAddEndPuncttrue
\mciteSetBstMidEndSepPunct{\mcitedefaultmidpunct}
{\mcitedefaultendpunct}{\mcitedefaultseppunct}\relax
\EndOfBibitem
\bibitem[Bhowmik and Ranganathan(2001)Bhowmik, and Ranganathan]{BHOWMIK200127}
Bhowmik,~R.; Ranganathan,~R. Cluster glass behaviour in
  Co0.2Zn0.8Fe2−xRhxO4(x=0–1.0). \emph{J. Magn. Magn. Mater.}
  \textbf{2001}, \emph{237}, 27--40\relax
\mciteBstWouldAddEndPuncttrue
\mciteSetBstMidEndSepPunct{\mcitedefaultmidpunct}
{\mcitedefaultendpunct}{\mcitedefaultseppunct}\relax
\EndOfBibitem
\bibitem[Padmanaban \latin{et~al.}(1990)Padmanaban, Avasthi, and
  Ghose]{PADMANABAN1990286}
Padmanaban,~N.; Avasthi,~B.; Ghose,~J. Solid state studies on
  rhodium-substituted CuCr$_2$O$_4$ spinel oxide. \emph{J. Solid State Chem.}
  \textbf{1990}, \emph{86}, 286--292\relax
\mciteBstWouldAddEndPuncttrue
\mciteSetBstMidEndSepPunct{\mcitedefaultmidpunct}
{\mcitedefaultendpunct}{\mcitedefaultseppunct}\relax
\EndOfBibitem
\bibitem[Weber \latin{et~al.}(2017)Weber, Schoop, Wurmbrand, Nuss, Seibel,
  Tafti, Ji, Cava, Dinnebier, and Lotsch]{weber2017}
Weber,~D.; Schoop,~L.~M.; Wurmbrand,~D.; Nuss,~J.; Seibel,~E.~M.; Tafti,~F.~F.;
  Ji,~H.; Cava,~R.~J.; Dinnebier,~R.~E.; Lotsch,~B.~V. Trivalent Iridium
  Oxides: Layered Triangular Lattice Iridate K$_{0.75}$Na$_{0.25}$IrO$_2$ and
  Oxyhydroxide IrOOH. \emph{Chem. Mater.} \textbf{2017}, \emph{29},
  8338--8345\relax
\mciteBstWouldAddEndPuncttrue
\mciteSetBstMidEndSepPunct{\mcitedefaultmidpunct}
{\mcitedefaultendpunct}{\mcitedefaultseppunct}\relax
\EndOfBibitem
\bibitem[Miao and Seshadri(2012)Miao, and Seshadri]{Miao_2012}
Miao,~M.-S.; Seshadri,~R. Rh$_2$O$_3$ versus IrO$_2$: relativistic effects and
  the stability of Ir$^{4+}$. \emph{J. Phys. Condens. Matter} \textbf{2012},
  \emph{24}, 215503\relax
\mciteBstWouldAddEndPuncttrue
\mciteSetBstMidEndSepPunct{\mcitedefaultmidpunct}
{\mcitedefaultendpunct}{\mcitedefaultseppunct}\relax
\EndOfBibitem
\bibitem[Roisnel and Rodr{\'{\i}}quez-Carvajal(2001)Roisnel, and
  Rodr{\'{\i}}quez-Carvajal]{roisnel2001}
Roisnel,~T.; Rodr{\'{\i}}quez-Carvajal,~J. WinPLOTR: A Windows Tool for Powder
  Diffraction Pattern Analysis. European Powder Diffraction EPDIC 7. 2001; pp
  118--123\relax
\mciteBstWouldAddEndPuncttrue
\mciteSetBstMidEndSepPunct{\mcitedefaultmidpunct}
{\mcitedefaultendpunct}{\mcitedefaultseppunct}\relax
\EndOfBibitem
\end{mcitethebibliography}
\end{document}